 \documentclass[preprints,review,accept,moreauthors,pdftex]{Definitions/mdpi} 

\makeatletter      
                    \renewcommand{\maketag@@@}[1]{\hbox{\m@th\normalsize\normalfont#1}}%
                    \makeatother
\firstpage{1} 
\pubvolume{xx}
\issuenum{1}
\articlenumber{5}
\pubyear{2019}
\copyrightyear{2019}
\history{Received: 25 October 2019; Accepted: 08 November 2019; Published: date}

\pdfoutput=1

\usepackage{amssymb}

\Title{Fast Vibrational Modes and Slow Heterogeneous Dynamics in Polymers and Viscous Liquids}

\Author{\hl{Francesco Puosi} $^{1}$\orcidA{}, Antonio Tripodo $^{1}$ and Dino Leporini $^{1,2,}$*\orcidB{}}
\AuthorNames{Francesco Puosi, Antonio Tripodo, and Dino Leporini}

\address{%
$^{1}$ \quad Dipartimento di Fisica ``Enrico Fermi'', Universit\`a di Pisa, Largo B. \@Pontecorvo 3, I-56127 Pisa, Italy; francesco.puosi@df.unipi.it (F.P.); antonio.tripodo@df.unipi.it (A.T.)\\
$^{2}$ \quad Istituto per i Processi Chimico-Fisici-Consiglio Nazionale delle Ricerche (IPCF-CNR), via G. Moruzzi 1, I-56124 Pisa, Italy}

\corres{Correspondence: dino.leporini@unipi.it; Tel.: +39-050-2214937}

\abstract{\textls[-15]{Many systems, including polymers and molecular liquids, when adequately cooled and/or compressed, solidify into a disordered solid, i.e., a glass. The~transition is not abrupt, featuring progressive decrease of the microscopic mobility and huge slowing down of the relaxation.} A~distinctive aspect of glass-forming materials is the microscopic dynamical heterogeneity (DH)
, i.e., the presence of regions with almost immobile particles coexisting with others where highly mobile ones are located. Following the first compelling evidence of a strong correlation between vibrational dynamics and ultraslow relaxation, we posed the question if the vibrational dynamics encodes predictive information on DH. Here, we review our results, drawn from molecular-dynamics numerical simulation of polymeric and molecular glass-formers, with a special focus on both the breakdown of the Stokes--Einstein relation between diffusion and viscosity, and the size of the regions with correlated displacements. 
}


\begin{document}


\section{Introduction}
\label{intro}
When polymers, liquids, biomaterials, metals and molten salts are cooled or compressed, if the crystallization is avoided,  they freeze into a microscopically disordered solid-like state, a glass~\cite{Angell91,DebeStilli2001,NgaiBook}. On approaching the glass transition from states with high fluidity, the viscosity exhibits a huge increase of more than 10 orders of magnitude~\cite{Angell91,DebeStilli2001}, along with the parallel decrease of the diffusivity~\cite{SILLESCURevDynHet99,NgaiBook}. Correspondingly, at microscopic level, solid-like behaviour becomes apparent, e.g., it is observed that a particle spends increasing time within the cage formed by the first neighbours where it rattles with amplitude $\langle u^2 \rangle^{1/2}$ on picosecond time scales~\cite{Ngai04}. This temporary trapping is rather persistent and  the particle has average escape time, the structural relaxation time $\tau_\alpha$, which increases from a few picoseconds in the low-viscosity liquid up to thousands of seconds close to the glass transition~\cite{BerthieRev}. 
{The~quantity  $\langle u^2 \rangle$  appears in the expression of the Debye--Waller (DW) factor, which,  assuming harmonicity and isotropy of the thermal motion, takes the form $\exp\left( -q^2 \langle u^2\rangle / 3 \right)$, where $q$ is the absolute value of the scattering vector~\cite{Ashcroft76}. Researchers investigating the cage motion in viscous liquids usually refer, as a metonym,  to $\langle u^2 \rangle$ as the DW factor too, e.g., see the work in~\cite{DouglasCiceroneSoftMatter12,DouglasStarrPNAS2015,DouglasLocalMod16}. To keep maintain similarity with this literature, the same convention is adopted here.
}

\begin{figure}[H]
\centering
\includegraphics[width=12 cm]{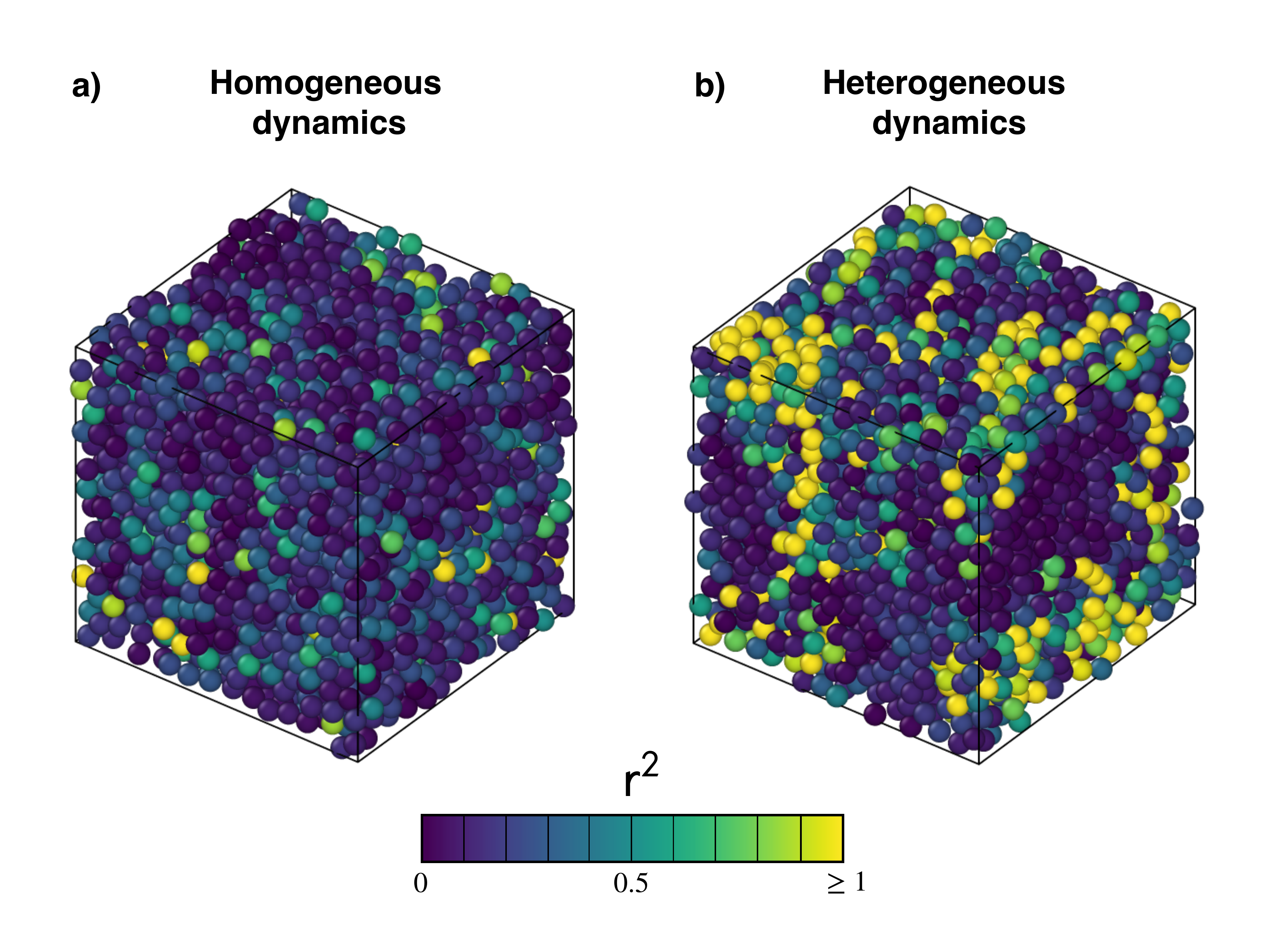}
\caption{Monomer arrangements at a time $t_0$ of two states of a polymer melt with (\textbf{a}) homogeneous and (\textbf{b}) heterogeneous dynamics. Bonds are removed for clarity reasons. Particles are coloured according to their squared displacements in the time interval $[t_0-\tau_\alpha, t_0]$. {Bright yellow particles have squared displacements no less than $1$. Notice that the two states have comparable mean square displacement ($\sim $0.21, homogeneous state; $\sim $0.28, heterogeneous state) but rather different relaxation times $\tau_\alpha$ ($\sim $9, homogeneous state; $\sim $1550, heterogeneous state).}
Homogeneous, i.e., position-independent, dynamics of the monomers is an aspect of systems with fast relaxation. Conversely, in the presence of heterogeneous dynamics, clusters of particles with extremely high mobility coexist with nearly immobile ones, slowing down the relaxation.}
\label{HOM_HET}
\end{figure}

The~transition from a liquid to a glass is accompanied by the growth of transient domains which exhibit different mobility, e.g., see Figure~\ref{HOM_HET}. The~phenomenon is usually dubbed ``dynamical heterogeneity'' (DH) and has been extensively studied, e.g., see the reviews in~\cite{SILLESCURevDynHet99,Ediger00,Richert02,BerthieRev} and topical papers~\cite{TarjusKivelsonJCP95,TrachtSpiessDynHetPRL98,DouglasLepoJNCS98,GarrahanStokesEinsteinChandlerPRE04}. The~size of the domains is relatively small involving approximately 10 molecule diameters~\cite{Ediger00}, corresponding to a few nanometres~\cite{TrachtSpiessDynHetPRL98}.  On a more general ground, the size of DH domains is strictly related to the possible presence of characteristic length scales in glass-forming systems. Starting with the seminal paper by Adam and Gibbs,  who invoked the presence  of ``cooperatively rearranging regions''  in viscous liquids~\cite{AdamGibbs65}, increasing interest has been devoted to identifying possible growing length scales as mobility decreases~\cite{StarrDouglasSastryDynHetJCP13,SastryLengthScalesRepProgrPhys15}. A broad classification in terms of either static or dynamic length scales is usually used. Static (thermodynamic) length scales are determined by the free-energy landscape, whereas dynamic length scales  are set by the rules governing the time evolution of the system and  extracted from finite-time behaviour of time-dependent correlation
functions and associated susceptibilities~\cite{BerthieRev}.  Even if growing static length scales have been reported by experiments~\cite{BiroliBouchaudLoidlLunkenheimerScience16} and simulations~\cite{BiroliKarmakarProcacciaPRL13}, there is still debate if they control the glass transition ~\cite{CatesWyartPRL17}. 
 It is still not clear to what extent dynamic correlations are the consequence, or the primary origin of, slow dynamics occurring close to the glass transition~\cite{SastryLengthScalesRepProgrPhys15}.
Even if rooted at nanometric length scales, DH exerts clear influence at macroscopic level. One~widely studied phenomenon is the breakdown of the Stokes--Einstein (SE) relation involving the diffusion coefficient $D$ and the shear viscosity $\eta$ (the more debated analogous phenomenon involving the rotational diffusion, where the breakdown is revealed~\cite{DiSchinoEPL97,CristianoDSE} or missing~\cite{Ediger00}, {will not be  considered here}).
For a single particle with radius $R$ moving in a homogeneous fluid with viscosity $\eta$ at temperature $T$, the SE relation states that
\begin{equation}
D = \frac{k_B T}{\zeta \pi \eta R}
\label{SE}
\end{equation}
$k_B$ denotes the Boltzmann constant and $\zeta$ denotes a number depending on the boundary condition between the fluid and the particle~\cite{Harris}. Under a no-slip condition, $\zeta = 6$. Roughly, the SE law states that the quantity  $k_BT/D\eta$ is a constant of the order of the size of the diffusing particle.  Remarkably, despite its macroscopic derivation, SE  also well accounts for the self-diffusion of many monoatomic and molecular liquids, provided the viscosity is low ($\lesssim$10  Pa $\cdot$s)~\cite{HansenMcDonaldIIIEd}. On the other hand,  the finite diffusion coefficient of guest atoms in solid hosts, where viscous transport is missing strongly, suggest the SE failure close to the solidification occurring at the glass transition. In fact, a common feature of several fragile glass formers is  the SE breakdown for increasing viscosity. The~failure manifests itself as a partial decoupling between the diffusion and the viscosity, in the sense that $D^{-1}$ increases less than $\eta$~\cite{SILLESCURevDynHet99,Ediger00,CristianoSE,LadJCP12,Puosi12SE,Puosi18SE}. The~decoupling is well accounted for by the fractional SE  $D\sim \eta^{-\kappa} $~\cite{SillescuSEJNCS94}, where the non-universal exponent $\kappa$ falls in the range $0.5 \le \kappa < 1$~\cite{DouglasLepoJNCS98}. 
The~usual interpretation of the SE breakdown relies on DH and the subsequent presence of a spatial distribution of characteristic relaxation times $\tau$ close to the glass transition~\cite{Ediger00,SillescuSEJNCS94,BerthieRev}. The~neat argument is that, although the viscosity is more sensitive to the longest relaxation times, the diffusivity is influenced by the shortest ones. As the shape of the distribution tends to widen on approaching the glass transition, the gap between $D^{-1}$ and $\eta$ increases as well, leading to the SE breakdown~\cite{SILLESCURevDynHet99}.

Diffusion, viscous transport, and structural relaxation involve time scales that are much longer that  than the typical vibrational time $t^\star$ of the particle rattling in the cage of the first neighbours, typically a few picoseconds. The~diffusion coefficient  is expressed as $D = 6 \delta^2/\tau_D$, where $\tau_D$ is { the minimum} time ensuring that the particle random displacements at a pace $\tau_D$ are statistically independent with finite mean square value $\delta^2$~\cite{Harris}. On the other hand, the viscous flow requires the relaxation of the shear stress fluctuation, which occurs in a Maxwell time $\tau_M = \eta/ G$, where $G$ is the intermediate-time shear modulus ~\cite{Puosi12}. On approaching the glass transition, $t^\star \ll \tau_D, \tau_M, \tau_\alpha$.

Despite the huge difference in time scales, earlier~\cite{TobolskyEtAl43} and later theoretical studies~\cite{HallWoly87,Angell95,Dyre96,MarAngell01,Sastry01,Ngai04,Ngai00,XiaWolynes00}, and experimental ones~\cite{BuchZorn92}, addressed the rattling process within the cage to understand the slow dynamics,  rising a growing interest on the DW factor
\cite{StarrEtAl02,BordatNgai04,Harrowell06,DouglasEtAlPNAS2009,Harrowell_NP08,OurNatPhys,lepoJCP09,Puosi11,UnivPhilMag11,OttochianLepoJNCS11,UnivSoftMatter11,Puosi12SE,PuosiLepoJCPCor12,PuosiLepoJCPCor12_Erratum,DouglasCiceroneSoftMatter12,SpecialIssueJCP13,CommentSoftMatter13,SokolovNovikovPRL13,TVG1,Merabia_JCP17,Vogel_JCP17,Puosi18SE,BecchiSoftMatter18}. Within this context, most interest has been devoted to the correlations between DW factor and the structural relaxation time $\tau_\alpha$, which are found to be strong and encompassed by a universal master curve~\cite{OurNatPhys}:
\begin{equation}
\tau_\alpha = \mathcal{F}(\langle u^2\rangle)
\label{eqn:u2tauExpGeneric}
\end{equation}

An analytical expression of the master curve is derived in Section~\ref{ModelDW}. Alternative forms of the master curve are reported by Douglas and coworkers~\cite{DouglasCiceroneSoftMatter12,DouglasStarrPNAS2015,DouglasLocalMod16}. Correlations between DW factor and the structural relaxation time $\tau_\alpha$ are found in polymers in bulk~\cite{OurNatPhys,lepoJCP09,Puosi11,Puosi18SE} and thin films~\cite{BecchiSoftMatter18}, binary atomic mixtures~\cite{lepoJCP09,SpecialIssueJCP13}, colloidal gels~\cite{UnivSoftMatter11}, antiplasticised polymers~\cite{DouglasCiceroneSoftMatter12,DouglasStarrPNAS2015}, water~\cite{SokolovNovikovPRL13} and water-like models~\mbox{\cite{Merabia_JCP17,Vogel_JCP17}}. The~DW factor also provided an alternative interpretation of the so-called thermodynamic (or temperature/density) scaling~\cite{TVG1}.
The~correlation between structural relaxation and DW factor has been inspected in the experimental data concerning several glass-formers in  a wide range of fragility---the steepness index $m$ defined by Angell~\cite{Angell91}  ($20 \le m \le 191$), including polymers, van der Waals and hydrogen-bonded liquids, metallic glasses, molten salts and the strongest inorganic glass-formers~\cite{OurNatPhys,UnivPhilMag11,OttochianLepoJNCS11,SpecialIssueJCP13,CommentSoftMatter13,SokolovNovikovPRL13,TVG1}. 

The~structural relaxation time $\tau_\alpha$ is an average quantity which is certainly affected by DH but not in a straightforward way. Nonetheless, given the scaling expressed by Equation~\eqref{eqn:u2tauExpGeneric}, it is legitimate to wonder if  DH and fast vibrational dynamics exhibit correlations. Working in that direction, we have found positive answers and the present paper collects and reviews a selected part of our results, with~a~focus on the breakdown of the SE law~\cite{Puosi12SE,Puosi18SE}. Even if strictly related, we will not discuss here a study concerning ultrathin molecular films with strong mobility gradients analogous to DH, where the same scaling observed in bulk, Equation~\eqref{eqn:u2tauExpGeneric}, has been revealed~\cite{BecchiSoftMatter18}.

\textls[-15]{Our approach relies on the increasing evidence that the master curve, Equation~\eqref{eqn:u2tauExpGeneric}, is a manifestation of a more fundamental correlation between the vibrational dynamics and  the slow relaxation.} It may be presented in the following terms. Let us consider a generic space- and time-dependent correlation function $C({\bf x}_1,t_1;  {\bf x}_2,t_2)$, where ${\bf x}$ denotes a configuration of the liquid at a given time $t$, i.e., the set ${\bf x}$ of all the positions of the elementary microscopic particles (monomers, atoms, molecules, etc.). For~steady states, $C({\bf x}_1,t_1;  {\bf x}_2,t_2)$ depends on the time difference $t = t_2-t_1$. Let us set $t_1=0$ and define $C({\bf x}_0; {\bf x}, t) \equiv C({\bf x}_0,0; {\bf x},t)$.  If two states,  labelled by $a$ and $b$, have equal DW factor, the correlation function, when evaluated over the two states, has coincident time evolution at least between the typical vibrational time scale $t^\star$ and $\tau_\alpha$~\cite{Puosi11}. Said otherwise, for $ t^* \lesssim t \lesssim \tau_\alpha$, it holds~\cite{OurNatPhys,Puosi11,UnivPhilMag11,OttochianLepoJNCS11,Puosi12,PuosiLepoJCPCor12,PuosiLepoJCPCor12_Erratum,SpecialIssueJCP13}:
\begin{equation}
\langle u^{2}\rangle_{(a)} = \langle u^2\rangle_{(b)} \hspace{1mm} \Rightarrow \hspace{1mm} C({\bf x}_0; {\bf x}, t)_{(a)} = C({\bf x}_0; {\bf x}, t)_{(b)}
\label{u2X}
\end{equation}

In selected systems, Equation~\eqref{u2X} holds beyond $\tau_\alpha$ and extends up to the diffusive regime, e.g., unentangled polymers and atomic binary mixtures ~\cite{Puosi11,SpecialIssueJCP13,Puosi12SE,Puosi18SE}.

Our studies were prompted by the finding by Widmer-Cooper and Harrowell that DH are predicted by particle displacements at short times ~\cite{Harrowell06}. However, it must be stressed that our DW factor is evaluated within the vibrational time scale $t^\star$ and not the time scale in~\cite{Harrowell06}, which is approximately one order of magnitude longer, a choice leading to differences for states with low viscosity. 

The~review outlines a model of the slow heterogeneous relaxation and transport in terms of vibrational dynamics in Section~\ref{ModelDW}. The~model is presented for completeness, but it is not essential to the understanding of the simulation results discussed in the rest of the paper. Later, a broad introduction to relaxation and transport in polymeric melts, and the correlation with the vibrational fast dynamics is given in Sections~\ref{relax} and  \ref{CorFastRelax}, respectively. The~signatures identifying the presence of heterogeneous dynamics are discussed in Section~\ref{HetDyn}. The~SE breakdown is presented in Section~\ref{SEbreak}, with a final discussion on the  length scale of the mutual influence between particle displacements in Section~\ref{DisplDisplCorFun}.

\section{\textls[-15]{A model of the Slow Heterogeneous Relaxation and Transport in Terms of Vibrational Dynamics}}
\label{ModelDW}

An in-depth, microscopic understanding of the link between the fast and slow dynamics is still missing, even if the impact of anharmonicity has been noted~\cite{BordatNgai04,TVG1,BerniniElasticJPCM17}. Here, we present a model, extending first seminal ideas~\cite{HallWoly87}, where the key role is played by the DW factor $\langle u^2 \rangle$, which is a single-particle quantity. Alternative pictures, in terms of the same quantity, are known ~\cite{DouglasCiceroneSoftMatter12,DouglasStarrPNAS2015,DouglasLocalMod16}. 
Notice that, even if  a single-particle quantity, the DW factor encodes information on collective dynamics and spatially extended cooperative phenomena~\cite{Puosi12,PuosiLepoJCPCor12,PuosiLepoJCPCor12_Erratum,ElasticoEPJE15,BerniniCageEffectJCP16,BerniniElasticJPCM17}.

At the present level of development, the model delivers expressions of the diffusion coefficient and the structural relaxation time in terms of the DW factor. It also accounts for the nonexponential character of the relaxation, an aspect which will be not presented here. However, even if it incorporates some consequence of DH, i.e., the presence of a wide distribution of relaxation times $p(\tau)$, it does not cover any spatial aspect related to DH, which instead has been revealed by the simulations, as we will see in Sections~\ref{VanHove} and  \ref{DisplDisplCorFun}, and accounted for by Equation~\eqref{u2X}.

\subsection{Relaxation Time} 
\label{relaxtime}
A first basis to connect fast and slow degrees of freedom was developed by Hall and Wolynes who, assuming that atomic motion is restricted to cells, pictured the glass transition as a freezing in an aperiodic crystal structure~\cite{HallWoly87}.
As a result, the viscous flow is described in  terms of  activated jumps over energy barriers $\Delta E \propto k_B T a^2/ \langle u^2\rangle$, where $a$ is the displacement to reach the transition state. The~usual rate theory leads to the Hall--Wolynes equation:
\begin{equation}
\label{dyreWolynes}
\tau_\alpha ^{(HW)} (a^2), \eta^{(HW)}(a^2) \propto \exp \left (\frac{a^2}{2\langle u^2\rangle}\right)
\end{equation}

Equation~\eqref{dyreWolynes} has the form of Equation~\eqref{eqn:u2tauExpGeneric}.
A very similar relation was derived by Buchenau and Zorn, in terms of soft vibrational modes~\cite{BuchZorn92}. \textls[-15]{Equation~\eqref{dyreWolynes} is expected to fail when $\tau_{\alpha}$ becomes comparable to the typical rattling times of each atom in the cage, corresponding to picosecond timescales. This condition is quite mild, e.g., in selenium it occurs \textasciitilde100 K above the melting temperature~\cite{BuchZorn92}. }

One basic assumption of  Equation~\eqref{dyreWolynes} is that the distance to reach the transition state  has a characteristic value $a$. Actually, this length scale is dispersed. To constrain the related distribution, $p(a^2)$, it is assumed that the latter does not depend on the state parameters such as the temperature, the density or the interacting potential. This complies with the spirit of the work in \cite{HallWoly87}, where the  $a$ distance is said to be mostly controlled by the geometrical packings. It is also known  that, irrespective of the relaxation 
time, $\tau_\alpha$, the average distance moved by the relaxing unit within $\tau_\alpha$ is approximately the
same, i.e., a~fraction of the molecular diameter~\cite{Angell91}. Averaging Equation~\eqref{dyreWolynes} over the distribution $p(a^2)$ yields the structural relaxation time
\begin{eqnarray}
\tau_\alpha  &=& \left \langle  \tau_\alpha^{(HW)}(a^2) \right \rangle_{a^2} \label{integral0} \\
&\equiv& \int_0^\infty  \tau_\alpha^{(HW)}(a^2) \,  p(a^2)d a^2
\label{integral}
\end{eqnarray}

Note that Equation~\eqref{integral} assumes that the distribution of the relaxation times is mainly due to the distribution of the displacement to reach the transition state in the different local environments, whereas the average DW factor $\langle u^2\rangle$ is taken as homogeneous across the sample. This viewpoint relies on the picture that relaxation is related to long wavelength soft modes~\cite{BuchZorn92,Harrowell_NP08}. Support has been provided by the strong correlation observed in glass-formers between $\langle u^2\rangle$ and the elastic modulus under quasi-static mechanical equilibrium~\cite{Puosi12}.

As a suitable choice, the distribution of the squared distances $p(a^2)$ is taken as  a truncated Gaussian form~\cite{OurNatPhys,lepoJCP09}
\begin{equation}
p(a^2) = \left\{ \begin{array}{ll}
A \exp\left (-\frac{(a^2 - \overline{a^2})^2}{2\sigma^2_{a^2}}\right )
 & \textrm{if $a>a_{min}$}\\
0& \textrm{otherwise}
\end{array} \right.
\label{Eq:GaussianDistro2}
\end{equation}
where $A$ is the normalization ensuring $\int_0^\infty p(a^2) \, d a^2 = 1$ and $a^2_{min}$ is the minimum displacement to reach the transition state. Given the weak influence, and to get rid of an adjustable parameter, one takes $a^2_{min}=0$~\cite{OurNatPhys,lepoJCP09}. The~motivations behind the Gaussian form of $p(a^2)$ mainly rely on the Central Limit Theorem. In fact, $a^2$ ($r^2_0$ in the notation in~\cite{HallWoly87}) is the cumulative squared displacement of the $N_{mon}$ particle that move~\cite{HallWoly87}.

Plugging Equation~\eqref{Eq:GaussianDistro2} into Equation~\eqref{integral} leads to the following generalized HW equation (GHW),
\begin{equation}
\tau_\alpha =  \tau_0 \;  \exp\left ( \frac{\overline{a^2}}{2\langle u^2\rangle } + \frac{ \sigma^2_{a^2}}{8\langle u^2\rangle ^2 } \right )
\label{parabola2}
\end{equation}
$\tau_0$ is a suitable constant. An analogous law is anticipated for the viscosity $\eta$, given the known near proportionality with $\tau_\alpha$~\cite{NgaiBook}. Equation~\eqref{parabola2} is the form of the master curve Equation~\eqref{eqn:u2tauExpGeneric} being adopted in our studies. Other variants useful in the comparison with numerical and experimental results are listed in Appendix \ref{app1} and \ref{appendixvanHove}.

Obviously, if the distribution $p(a^2)$ is narrow {and centred at $a^2_0$, Equation~\eqref{parabola2} must reduce to the expression derived by Hall and Wolynes, Equation~\eqref{dyreWolynes}, $\tau_\alpha ^{(HW)} (a^2_0)$}.  For the specific choice of  $p(a^2)$, given by Equation~\eqref{Eq:GaussianDistro}, Equation~\eqref{parabola2} shows that this happens if  $\sigma^2_{a^2}/8\langle u^2\rangle ^2 \ll \overline{a^2}/2\langle u^2\rangle$, namely,  the ratio $R$ defined as
\begin{equation}
R \equiv \sigma^2_{a^2}/4 \overline{a^2} \langle u^2\rangle
\label{DefR2}
\end{equation}
is vanishingly small. Equation~\eqref{DefR2} depends on the magnitude of DW factor so that, being the parameters $\sigma^2_{a^2}$ and $\overline{a^2}$ independent of the physical state, the presence of the distribution $p(a^2)$ is negligible when the DW factor is large, thus leading to a very narrow distribution of relaxation times, a characteristic of homogeneous dynamics. This suggests to read the condition $R=1$ as the crossover between homogeneous and heterogeneous dynamics, i.e.,\vspace{12pt}
\begin{equation}
\left\{ \begin{array}{ll}
R \ll 1
 & \textrm{homogeneous dynamics}\\
R \gg 1& \textrm{heterogeneous dynamics}
\end{array} \right.
\label{Eq:GaussianDistro}
\end{equation}

Finally, we notice that the distribution $p(a^2)$ in Equation~\eqref{Eq:GaussianDistro2} with $a^2_{min}=0$ may be recast via Equation~\eqref{dyreWolynes}, as a log-normal distribution of relaxation times $p(\ln\tau)$
\begin{equation}
p(\ln\tau) =   \left\{ \begin{array}{ll}
B \exp \left \{ - \frac{2\langle u^2\rangle^2}{\sigma^2_{a^2}} \left [ \ln \left ( \frac{\tau}{\overline{\tau}} \right ) \right ]^2   \right \}
 & \textrm{if  $\tau \ge \tau_0$}\\
0& \textrm{otherwise}
\end{array} \right.
\label{lognormal}
\end{equation}
where $B$ is the normalization ensuring $ \int p(\ln\tau) \, d \ln \tau = 1$, $\overline{\tau} = \tau_0 \exp (\overline{a^2}/2\langle u^2\rangle )$. An interesting feature of $p(\ln\tau)$ is that its width $\sim \sigma_{a^2}/ \langle u^2\rangle$ increases by decreasing the DW factor.

\subsection{Diffusion Coefficient}
\label{diffcoeff}
The~diffusion coefficient $D$ may be expressed via the above model  by the  relation~\cite{Puosi18SE}
\begin{equation}
D =   \frac{1}{6} \left \langle \frac{a^2}{\tau_\alpha^{(HW)}(a^2)} \right \rangle_{a^2}
\label{diffusion}
\end{equation}

The~above equation assumes that displacements as large as $a$ occurring in a time $\tau_\alpha^{(HW)}(a^2)$ are statistically independent. 
Notice that, although Equation~\eqref{integral} signals that the structural relaxation time is affected by the larger $a^2$ values, i.e., the longest relaxation times of the distribution $p(\ln\tau)$, the diffusivity, according to Equation~\eqref{diffusion}, is influenced by the shorter ones.

The~explicit expression of the diffusion coefficient and an approximated version are given in Appendix \ref{appendixDiff}.

\subsection{Stokes--Einstein Product}
\label{SecSE}
The~Stokes--Einstein (SE) relation, Equation~\eqref{SE},  states that the quantity $D\eta/T$ is constant if the diffusing particle changes neither the size nor  the boundary conditions with the liquid. As the numerical evaluation of the viscosity is a delicate point,  proxies are often used~\cite{CristianoSE,allentildesley}. As an example, as  $\eta \propto T \tau_\alpha$ in unentangled polymers~\cite{DoiEdwards},  it is more suitable to study the breakdown of the SE law by considering the SE product
\begin{equation}
K_{SE} = D M \tau_{\alpha}
\label{SEProduct}
\end{equation}
where $M$ is the number of monomers. $K_{SE}$ is expected to be independent of the chain length, as $D \propto 1/M$ in unentangled polymers~\cite{DoiEdwards} and the monomer relaxation at $\tau_{\alpha}$ poorly senses the chain connectivity. The~above equation with $M=1$ may be also used for liquids where the elementary units are atoms or small molecules, as the temperature factor in the ratio $D\eta/T$ provides a change of approximately $\sim $20\% in fragile glass-formers~\cite{NgaiBook}, much less than the observed increase of $K_{SE}$ on approaching the glass transition~\cite{SILLESCURevDynHet99,Ediger00}. The~explicit expression of the SE product $K_{SE} $ derived within the vibrational dynamics model  and an approximated version $\tilde{K}_{SE}$ are given in Appendix~\ref{appendixSE}.

\section{Transport and Relaxation in Polymeric Melts}
\label{relax}

The~correlation between diffusivity, slow relaxation and fast vibrational dynamics has been studied by Molecular-Dynamics (MD) simulations of a coarse-grained model of a melt of linear unentangled polymer. Details about the model are given in Section~\ref{method}. Even if rather crude, the model was proven to capture the universal aspects of the correlation and allowed an effective comparison with the experiment~\cite{OurNatPhys}. 

To provide a microscopic picture of the transport, the mean square displacement (MSD) of the monomer $\langle r^2(t)\rangle$ is usually considered:\vspace{12pt}
\begin{equation}\small
\langle r^2(t)\rangle = \frac{1}{N} \sum_i \langle \|{\bf x}_i(t) - {\bf x}_i(0)\|^2  \rangle
\label{Eq:MSD}  
\end{equation}
where ${\bf x}_i(t)$ is the position of the $i$-th monomer at time $t$. In addition to MSD, with the purpose of characterizing the relaxation, the self part of the intermediate scattering function (ISF) is also considered~\cite{HansenMcDonaldIIIEd}:
\begin{equation}\small
F_{s}(q,t) = \frac{1}{N} \langle \sum_j^N e^{i{\bf q}\cdot({\bf x}_j(t) - {\bf x}_j(0))} 
\label{Eq:Fself}
\rangle 
\end{equation}

 {In an isotropic liquid, ISF 
 	 depends only on the modulus of the wavevector $q = || {\bf q} ||$ and features the rearrangements of the spatial structure of the fluid over the length scale $\sim$$2\pi/q$, leading to a decaying profile in time starting from $F_{s}(q,0)=1$.  In our case,
 ISF was evaluated at $q= q_{max}$, the~maximum of the static structure factor  ($7.13 \le q_{max}\le 7.55$) corresponding to the length scale of the monomer size. $F_{s}(q_{max},t)$ vanishes when the monomer displacement in a time $t$ largely exceeds the monomer diameter}. The~time needed to make $F_{s}(q_{max},t)$ small is a measure of the escape time of the monomer from the cage formed by the neighbours, also known as the structural relaxation time $\tau_\alpha$, customarily defined by the relation $F_s(q_{max}, \tau_{\alpha}) = e^{-1}$.

\begin{figure}[H]
\centering
\includegraphics[width=13 cm]{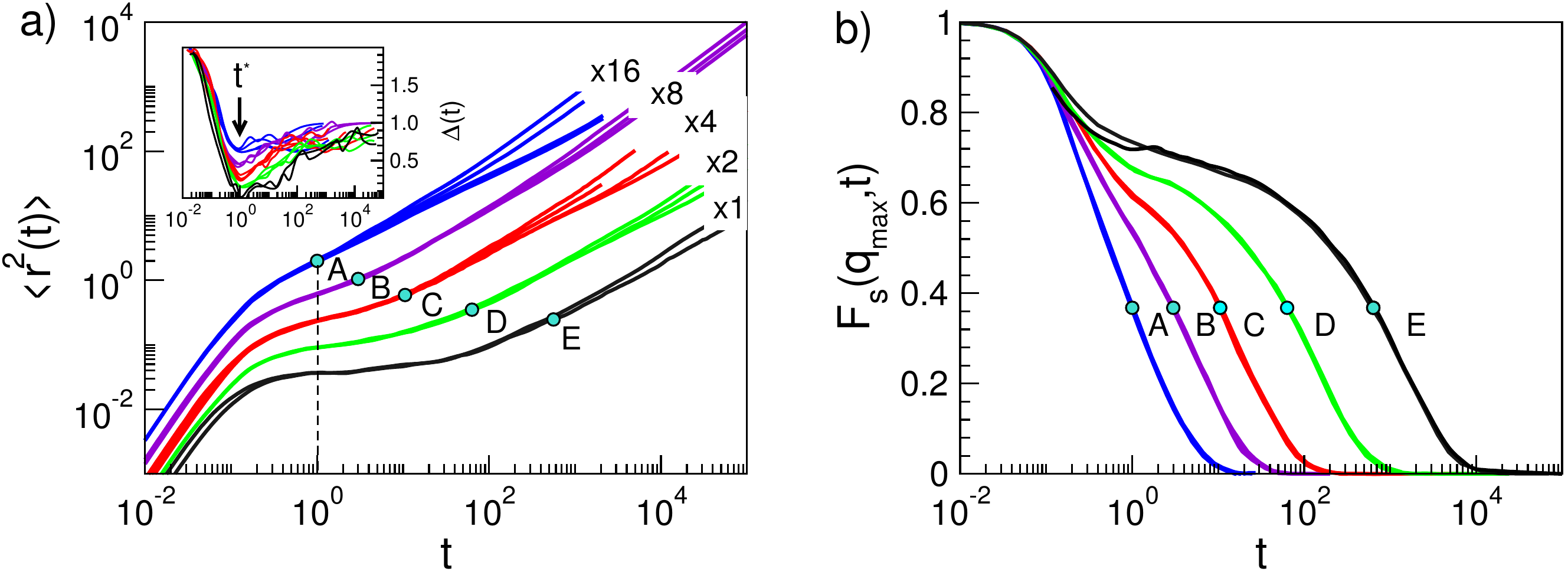}
\caption{Monomer dynamics in the polymer melt. (\textbf{a}) Mean square displacement (MSD) for polymers in selected states (see below for details). For clarity reasons, MSDs are multiplied by indicated factors. Inset: corresponding MSD slope $\Delta(t)$, Equation~\eqref{delta}; the position of the minimum at $t^{\star} = 1.0(4)$ is signalled by the arrow in the inset and the dashed line in the main panel.  (\textbf{b}) corresponding ISF curves. The~figure shows a manifestation of Equation~\eqref{u2X}{, see Section~\ref{VanHove} for details}, i.e., if states have equal DW factor $\langle u^2 \rangle$, both the MSD and ISF curves coincide at least in the time window $[t^{\star}, \tau_{\alpha}]$ ($\tau_{\alpha}$ is marked with dots on each curve).  The~physical states are labelled by the string (M, $\rho$, T, q, p) where $M$ is the number of monomers per chain, $\rho$ the number density, $T$ the temperature and the pair $(q,p)$ refers to the characteristic parameters of the non-bonding potential, Equation~\eqref{Eq:modifiedLJ}. The~six sets of states are as follows.
Set~A: (2,1.086,0.7,7,6), (3,1.086,0.7,7,6), (10,1.086,0.7,7,6), (10,1.033,0.7,8,6).  
Set~B: (2,1.033,0.7,10,6), (3,1.039,0.7,11,6), (3,1.041,0.7,11,6).
Set~C: (2,1.033,0.5,10,6), (3,1.056,0.7,12,6), (5,1.033,0.6,12,6), (10,1.056,0.7,12,6).
Set D: (3,1.086,0.7,12,6), (5,1.086,0.7,12,6), (10,1.086,0.7,12,6).
Set E: (2,1.0,0.7,12,11), (3,1.1,1.1,15,7). Data from ~\cite{lepoJCP09}.}
\label{fig1JCP09}
\end{figure}

Figure~\ref{fig1JCP09} shows typical MSD and ISF curves of the polymeric monomers. At very short times (ballistic regime), MSD increases according to $\langle r^2(t)\rangle \cong (3 k_B T/m) t^2$ and ISF starts to decay. The~repeated collisions with the other monomers slow the displacement of the tagged one, as evinced by the knee of MSD at $ t \sim \sqrt{12}/\Omega_0 \sim 0.17$, where $\Omega_0$ is an effective collision frequency, i.e., it is  the mean small oscillation frequency of the monomer in the potential well produced by the surrounding ones kept at their equilibrium positions~\cite{Boon,BerniniCageEffectJCP16}.
At later times, a quasi-plateau region, also found in ISF, occurs when the temperature is lowered and/or the density increased. This signals the increased caging of the particle. Trapping is terminated after an average time $\tau_{\alpha}$. For $t \gtrsim \tau_{\alpha}$, MSD increases more steeply. The~monomers of short chains ($M \lesssim 3$) undergo diffusive motion $\langle r^2(t)\rangle \propto t^\delta$ with $\delta=1$. For longer chains,  owing to the increased connectivity,  the onset of the diffusion is preceded by a subdiffusive region ($\delta < 1$, Rouse regime)~\cite{Gedde95}. At long time, the monomer displaces in a diffusive way with diffusion coefficient $D = \lim_{t \to \infty} \langle r^2(t)\rangle/6t$.

\section{Correlation between Vibrational Fast Dynamics and Slow Relaxation}
\label{CorFastRelax}
\vspace{-6pt}

\subsection{Vibrational Caged Dynamics and Debye--Waller factor}

In our model polymer, the term ``vibrational dynamics'' refers to the rattling of the trapped monomer within the cage formed by the closest monomers. It is crucial to provide a robust criterion to assess the presence of the cage, which is anticipated to lack in liquids with high molecular mobility and fast relaxation. Compelling evidence of the cage effect is provided by the time velocity correlation function, which, after a first large drop due to pair collisions, reverses the sign since the monomer rebounds from the cage wall~\cite{BerniniCageEffectJCP16}. As an alternative route to reveal the cage effect, we consider the slope of MSD in the log-log plot
\begin{equation}
\Delta(t) \equiv \frac{\partial \log \langle r^2(t)\rangle}{\partial \log t}
\label{delta}
\end{equation}

Representative plots of $\Delta(t)$ for the polymer system are given in the inset of Figure~\ref{fig1JCP09}{a}. $\Delta(t)$ tends to 2 at short times, due to the ballistic motion, and reaches the plateau level 1 at long times, owing to the diffusive motion. In the absence of caging effect, $\Delta(t) $ decreases in a monotonous way on increasing time. Caging is indicated by the presence of a minimum of $\Delta(t)$ occurring, irrespective of the physical state in the present model polymer,  at $t^{\star} = 1.0(4)$. In actual time units, $t^{\star}$ is \textasciitilde1--10 ps~\cite{Kroger04}.

The~presence of the minimum paves the way to a robust definition of the DW factor $\langle u^2\rangle$, the~mean square rattling amplitude of the monomer during the trapping period. In fact, the minimum, corresponding to the inflection point in the log-log plot of $\langle r^2(t)\rangle$), separates two regimes. At short times, $t  <  t^{\star}$, the inertial effects dominate, whereas for $t> t^{\star}$, early escapes from the cage become apparent. Therefore, a convenient definition of the DW factor as a mean localization length is just MSD~at~$t^{\star}$:
\begin{equation}
\langle u^2\rangle \equiv  \langle r^2(t=t^{\star})\rangle
\label{DWdef}
\end{equation}

\subsection{Debye--Waller Scaling of the Slow Relaxation}

The~monomer dynamics depends in a complex way on the state parameters. 
Nonetheless, there is clear correlation between the DW factors and the long-time relaxation dynamics. First examples are shown in Figure~\ref{fig1JCP09} by considering MSD and ISF. { Note that states with equal DW factor have coincident time evolution of both MSD and ISF at least between $t^\star$ and $\tau_\alpha$~\cite{Puosi11}.  In Section~\ref{VanHove}, it will be shown that these results are a manifestation of Equation~\eqref{u2X}.}

\begin{figure}[H]
\centering
\includegraphics[width=8 cm]{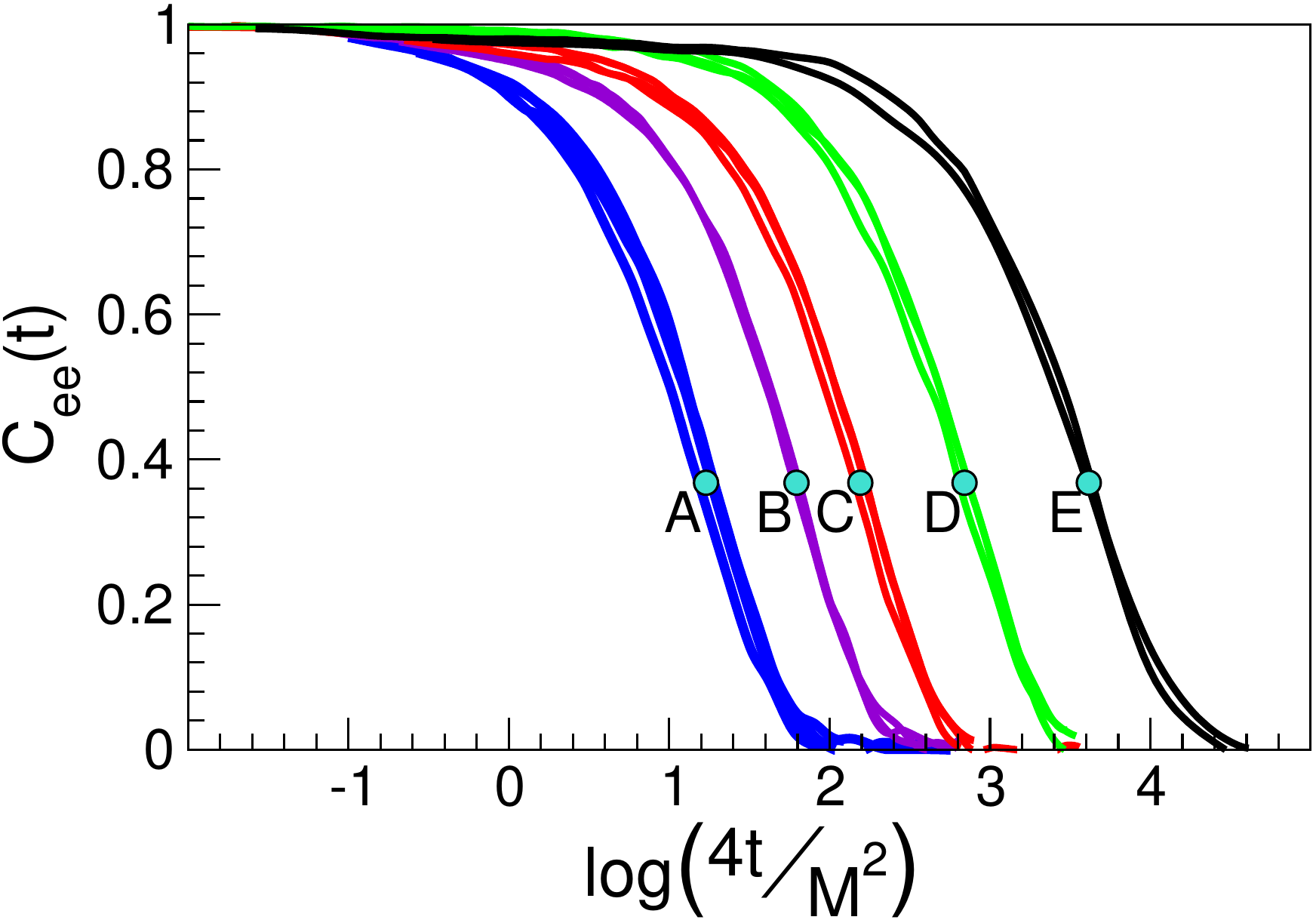}
\caption{Correlation function of the end-to-end vector joining the two ends of a polymer chain. Each group of curves corresponds the physical states A, ${\dots}$, E  with identical DW factor detailed in Figure~\ref{fig1JCP09}. Polymer states contributing to one cluster of scaled curves have not necessarily equal chain length. However, the scaled time removes the chain length dependence.  Dots mark the time $4 \tau_{ee}/ M^2$. The~results prove that Equation~\eqref{u2X} holds also at times $\tau_{ee}$ much longer than $\tau_\alpha$. Data  from~\cite{lepoJCP09}.}
\label{fig3JCP09}
\end{figure}

It is seen that the coincidence of the MSD curves is lacking at times longer than $\tau_\alpha$ for states corresponding to polymer chains with different length. This effect is not a failure of the scaling at times exceeding $\tau_\alpha$, but a mere consequence of the complex dependence of MSD on the chain length since it is affected by all the Rouse modes~\cite{DoiEdwards}. { In fact, if the correlation function of the single Rouse mode with the slowest relaxation time is singled out, i.e., the one with characteristic relaxation time given by the average chain reorientation time $\tau_{ee}$~\cite{DoiEdwards}, the scaling is still observed after proper account of the chain length dependence, see Figure~\ref{fig3JCP09}. The~finding proves that Equation~\eqref{u2X} holds also at a time $\tau_{ee}$ being much longer than $\tau_\alpha$}.
   
\begin{figure}[H]
\centering
\includegraphics[width=8 cm]{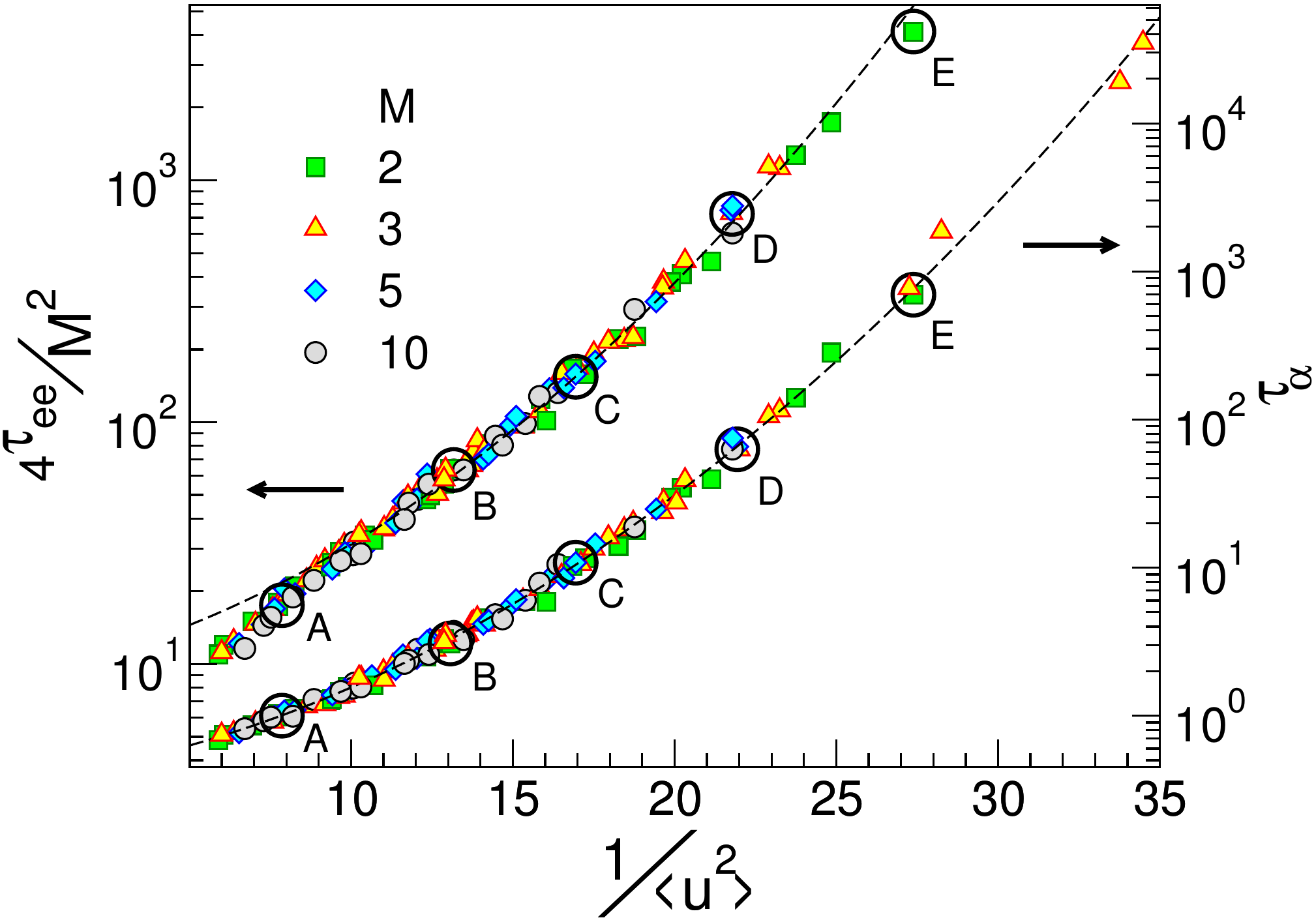}
\caption{The~structural relaxation time $\tau_\alpha$ and the scaled reorientation time $\tau_{ee}$ of the polymer chains vs. the DW factor $\langle u^2\rangle$. Empty circles highlight the cases plotted in Figure~\ref{fig1JCP09}. The~other states are detailed in Ref.~\cite{lepoJCP09}. The~dashed line across the $\tau_\alpha$ curve is Equation~\eqref{parabolaMD}. The~dashed curve  across the chain reorientation time  curve is a guide for the eyes. Data from~\cite{lepoJCP09}.}
\label{fig7JCP09}
\end{figure}

As a side product of the coincidence of the ISF curves in states with equal DW factor seen in Figure~\ref{fig1JCP09}, one has that states with equal DW factor $\langle u^2\rangle$ have equal structural relaxation time  $\tau_\alpha$ too. This can be reformulated via the master curve Equation~\eqref{eqn:u2tauExpGeneric}, which, according to the model detailed in Section~\ref{relaxtime}, takes the form given by Equation~\eqref{parabola2}, i.e., a simple parabolic law between $\log \tau_\alpha$ and $1/\langle u^2\rangle$~\cite{OurNatPhys,lepoJCP09}. Figure~\ref{fig7JCP09} tests Equation~\eqref{parabola2}, written in the form given by Equation~\eqref{parabolaMD} for a wide variety of physical states of our model polymeric melt~\cite{lepoJCP09}. It is also shown that the scaling holds if one considers the end--end chain reorientation time $\tau_{ee}$, i.e., the time needed by the correlation function $C_{ee}(t)$ to drop to $e^{-1}$, see Figure~\ref{fig3JCP09}; although, in this case, it is described by a different master curve.  
 \unskip

\section{Signatures of the Heterogeneous Dynamics}
\label{HetDyn}

MSD and ISF well-expose the cage effect, whereas the possible DH influence on their shape is less apparent.  Figure~\ref{HOM_HET} shows that DH is characterized by the presence of clusters of monomers with rather different mobility~\cite{Ediger00,Richert02}. We now present and discuss two quantities well tailored to provide quantitative insight into this aspect.

\subsection{van Hove Function}
\label{VanHove}
One central quantity of the DH analysis is the self part of the van Hove function 
$G_{s}({\bf r},t)$~\cite{HansenMcDonaldIIIEd}:
\begin{equation}
G_{s}({\bf r},t) = \frac{1}{N} \langle \sum_{i=1}^N  \delta [{\bf r}+{\bf x}_i(0)-{\bf x}_i(t)] \rangle
\label{vanhove}
\end{equation}
\noindent where ${\bf x} _i(t)$ is the position of the \emph{i}-th monomer at time $t$, and $\delta[\cdot] $ is the three-dimensional Dirac delta function. In isotropic liquids, the van Hove function depends on the 
modulus $r$ of ${\bf r}$. The~interpretation of $G_{s}(r,t)$ is 
direct. The~product $G_{s}(r,t) \cdot 4\pi r^{2}$ is the 
probability that the monomer is at a distance between $r$ and 
$r+dr$ from  the initial position after a time t. 
The~moments of $G_{s}(r,t)$ are of interest: \vspace{12pt}
\begin{equation}
\langle r^{n}(t)\rangle = 4 \pi \int_0^\infty r^{n} G_{s}(r,t) \; r^2 d r
\label{Eq:MSD}  
\end{equation}

For $n =2$, one recovers the usual mean square displacement (MSD). If the monomer displacement is a Gaussian random variable,
$G_{s}(r,t)$  reduces to the Gaussian form~\cite{HansenMcDonaldIIIEd}:
\begin{equation}
G^g_s(r,t)=\left(\frac{3}{2\pi\langle r^2(t)\rangle}\right)^{3/2}\exp\left(-\frac{3r^2}{2\langle r^2(t)\rangle}\right)
\label{vanhovegauss}
\end{equation}

Equation~\eqref{vanhovegauss} is the correct limit of $G_{s}(r,t)$ at very
short  (ballistic regime, $\langle r^2(t)\rangle = 3k_B T/\mu t^{2}$) 
and very long times ( diffusion regime, $\langle r^2(t)\rangle = 6 D t$, where $D$ is the monomer diffusion coefficient).

The~spatial Fourier transform of  the self part of the van Hove function yields the ISF function, Equation~\eqref{Eq:Fself}~\cite{HansenMcDonaldIIIEd}.

\begin{figure}[H]
\centering
\includegraphics[width=14 cm]{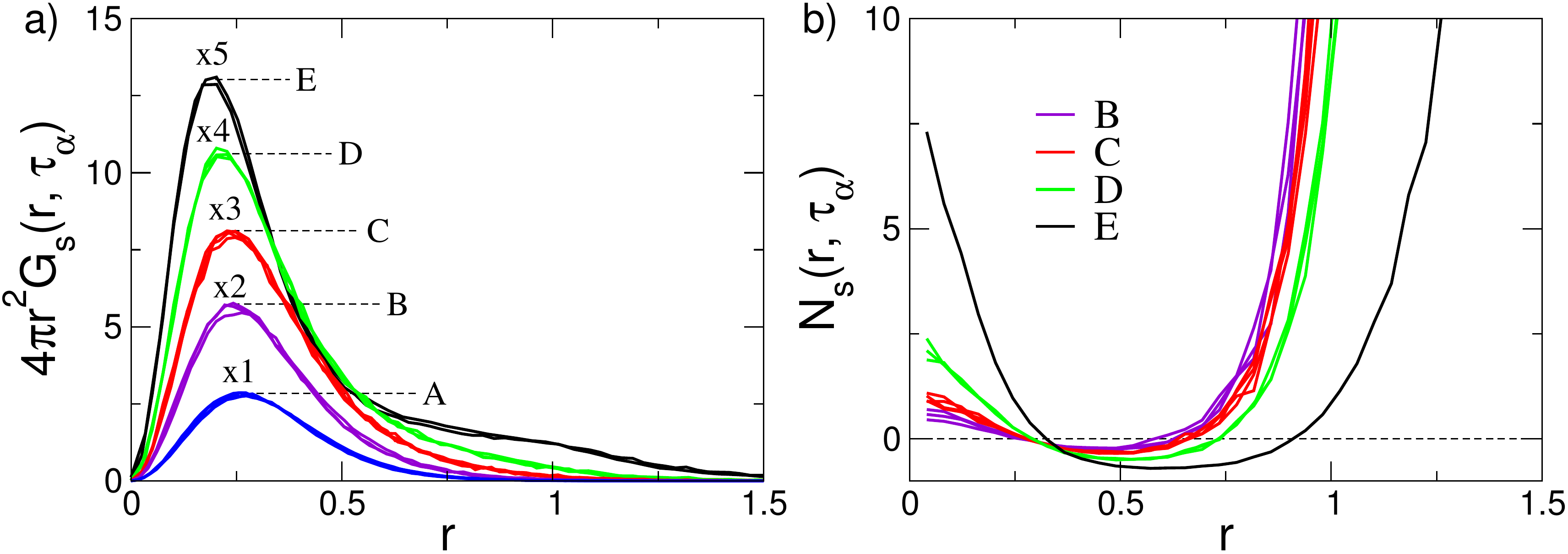}
\caption{(\textbf{a}) Self part of the van Hove function $G_s(r,t)$ of the states of  Figure~\ref{fig1JCP09} at the structural relaxation time $t=\tau_\alpha$. The~curves are multiplied by indicated factors. The~sets of clustered curves A--E show that, if states have equal DW factor, they have coincident van Hove functions too. { As $G_s(r,t)$ may be expressed in terms of correlation functions, the coincidence reflects Equation~\eqref{u2X}.} Data from~\cite{Puosi11}. 
(\textbf{b}) The~ratio $N_s(r,\tau_\alpha)$, Equation~\eqref{N}, of the states of Figure~\ref{fig1JCP09}.  On increasing the structural relaxation time from A states to E states, the system tends to increase the fractions of monomers with either much lower or much higher mobility with respect to the fraction predicted by the Gaussian approximation. 
Data from ~\cite{PuosiLepoJCPCor12}.
}
\label{fig4JPCB11fig5JCP12}
\end{figure}

Figure~\ref{fig4JPCB11fig5JCP12}a presents the self-part of the van Hove function $G_{s}(r,t)$, evaluated  at $\tau_\alpha$ for the set of states with different mobility and relaxation shown in Figure~\ref{fig1JCP09}. It is seen that if the relaxation and the mobility are fast, the shape of $G_{s}(r,\tau_\alpha)$ decreases by increasing the displacement $r$ from the initial position. On the other hand, the states belonging to the D and E set, the ones with slowest relaxation, exhibit a tendency toward a  bimodal pattern, namely, in addiction to particles undergoing small displacements, a shoulder at $r \sim 1$ (the monomer diameter) is observed. This signals the presence of particles exhibiting fast displacements by solid-state jump dynamics~\cite{CristianoSE}. Said otherwise, the quasi-bimodal pattern of the van Hove function is clear signature of DH. Four other aspects are to be noted: 
\begin{itemize}[leftmargin=*,labelsep=5mm]
\item{The self-part of the van Hove function is expressed by suitable correlation functions, see Appendix~\ref{appendixvanHove}. Then, the coincidence of $G_s(r,\tau_\alpha)$ in states with equal DW factor observed in Figure~\ref{fig4JPCB11fig5JCP12}a (the sets of states labelled as A, $\dots$, E) is in harmony with Equation~\eqref{u2X}. 
\item Equation~\eqref{u2X} also holds if one inspects the spatial dependence of the correlation function, e.g., the~van~Hove function, at $\tau_\alpha$. In particular, even in the presence of DH.
\item Given their relation with  $G_{s}(r,t)$, the coincidence of both MSD and ISF observed in Figure~\ref{fig1JCP09} for the sets of states labelled as A, $\dots$, E is strictly linked to the one observed in Figure~\ref{fig4JPCB11fig5JCP12}a.}
\item The pattern of the D and E sets of states is not consistent with the Gaussian limit $G^g_s(r,\tau_\alpha)$, Equation~\eqref{vanhovegauss}, predicting a progressive decay with $r$, i.e., the DH dynamics is {\it not} Gaussian; 
\end{itemize}

To quantify the deviations of the self-part of the van Hove function $G_{s}(r,\tau_\alpha)$ from the Gaussian limit, one defines the quantity~\cite{KobDonatiPRL97,CristianoSE}
\begin{equation}
N_s(r,\tau_\alpha)=\frac{G_s(r,\tau_\alpha)-G_s^g(r,\tau_\alpha)}{G_s^g(r,\tau_\alpha)}
\label{N}
\end{equation}

Figure~\ref{fig4JPCB11fig5JCP12}b plots the ratio $N_s(r,\tau_\alpha)$. It exhibits increasing positive deviations at both short and large $r$ values, evidencing the excess of nearly immobile and highly mobile monomers with respect to purely Gaussian behaviour, respectively.  
The~analysis, in terms of the ratio $N_s(r,\tau_\alpha)$, reveals the wide distribution of mobilities pictured in Figure~\ref{HOM_HET}, right. 

\subsection{Non-Gaussian Parameter}
\label{nongaspar}
An effective quantity to expose the time evolution of the non-Gaussian character of DH dynamics is the non-Gaussian parameter (NGP)~\cite{HansenMcDonaldIIIEd}:
\begin{equation}
\alpha_2(t) =  \frac{3}{5} \frac{\langle r^4(t)\rangle}{\langle r^2(t)\rangle^2} - 1 
\label{Eq:alpha2}
\end{equation}
where $\langle r^2(t) \rangle$ and $\langle r^4(t) \rangle$ are the mean square and quartic displacements of the particle at time $t$, respectively. $\alpha_2$ vanishes if the displacement is Gaussian, i.e., it follows from a series of independent elementary steps with finite mean and variance.

\begin{figure}[H]
\centering
\includegraphics[width=8 cm]{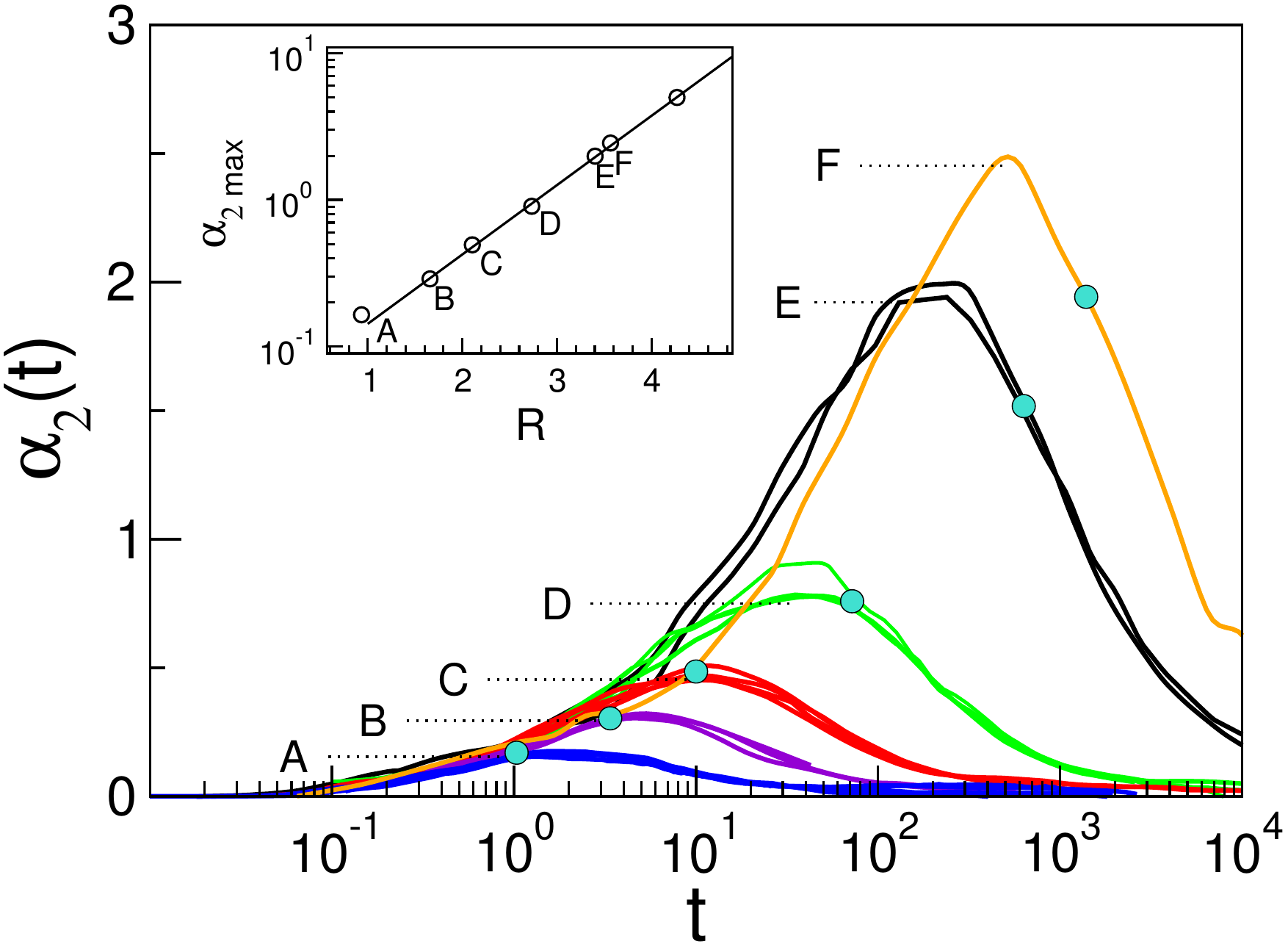}
\caption{Non-Gaussian parameters (NGPs) of states with different relaxation times $\tau_{\alpha}$ (marked with grey dots).  The~physical states A, ${\dots}$, E  are the states with identical DW factor detailed in Figure~\ref{fig1JCP09}. Note that they have coinciding NGPs in the time window $[t^\star, \tau_\alpha]$ at least, in agreement with Equation~\eqref{u2X}. The~curve labelled as F is the state (M, $\rho$, T, q, p) =  (3, 1.1, 0.65, 12, 6) with $\tau_\alpha \simeq 2  \cdot 10^3$, see Figure~\ref{fig7JCP09}.  
Inset: the NGP maximum $\alpha_{2}^{max}$ vs. the ratio $R$, Equation~\eqref{DefR2}. The~dot with the largest $\alpha_{2}^{max}$ value corresponds to the state with the longest structural relaxation time $\tau_\alpha$ in Figure~\ref{fig7JCP09} with parameters (M, $\rho$, T, q, p) = (3, 1.2, 0.95, 6,12). Data from~\cite{Puosi12SE}.}
\label{fig2JCP12SE}
\end{figure}

Figure~\ref{fig2JCP12SE} plots  the NGP time evolution, Equation~\eqref{Eq:alpha2}, for the set of states A, $\cdots$, E and additional states with very slow relaxation. It is seen that NGP vanishes at very short times, as the ballistic regime is Gaussian in nature. At intermediate times, a peak value $\alpha_2^{\, max}$  is observed increasing with the relaxation times~\cite{sim,BennemannNat99,Harrowell_NP08}. The~maximum 
occurs at a time slightly shorter than the structural relaxation time $\tau_\alpha$, as in simpler molecular systems~\cite{CristianoSE}. A snapshot of the microscopic mobilities in a lapse of time $\tau_\alpha$, where DH is quite apparent, is plotted in Figure~\ref{HOM_HET} (right). At later times, NGP decreases as the monomer dynamics enters the homogeneous diffusive regime, which is a Gaussian process~\cite{Harris}. 

 It is seen that states belonging to the same set A, $\cdots$, E, i.e., with equal DW factor, have identical NGP in the time window $[t^\star, \tau_\alpha]$ at least. { This agrees with Equation~\eqref{u2X},  given the relation of NGP with the moments of the self part of the van Hove function $G_s(r,t)$, Equation~\eqref{Eq:MSD}, and the expression of the latter in terms of suitable correlation functions, see Appendix \ref{appendixvanHove}}~\cite{NGP_Wolynes}. Note also the exponential increase of $\alpha_2^{\, max}$ with the ratio $R$ defined in Equation~\eqref{DefR2}~\cite{OurNatPhys,lepoJCP09}. This is in harmony with the inequalities in Equation~\eqref{Eq:GaussianDistro}, stating that DH is characterized by $R > 1$.

\section{Breakdown of the Stokes--Einstein (SE) Law in the Presence of Dynamical Heterogeneity}
\label{SEbreak}

\subsection{SE Breakdown in Unentangled Polymers}
\label{SEpolyme}

The~SE law is usually derived by considering the diffusivity of macroscopic bodies displacing in homogeneous viscous liquids~\cite{Harris}.  The~diffusion in the presence of strong DH  does not comply with the SE law~\cite{DouglasLepoJNCS98}. We have studied the SE breakdown in melts of unentangled polymers~\cite{Puosi12SE}. In~these systems, helpful features are found~\cite{DoiEdwards}: (i) the diffusion coefficient $D$ is inversely proportional to the chain length $M$, and (ii) the viscosity $\eta$ is proportional to the end--end reorientation time which, in~turn, is proportional to the structural relaxation time, e.g., see Figure~\ref{fig3JCP09}, showing that states with equal structural relaxation time also have equal end--end reorientation time. Then, as discussed in Section~\ref{SecSE}, the study of the validity of the SE law is more efficiently carried out in terms of the product $D M \tau_{\alpha}$, which is anticipated to be state-independent if the SE law holds.

\begin{figure}[H]
\centering
\includegraphics[width=14 cm]{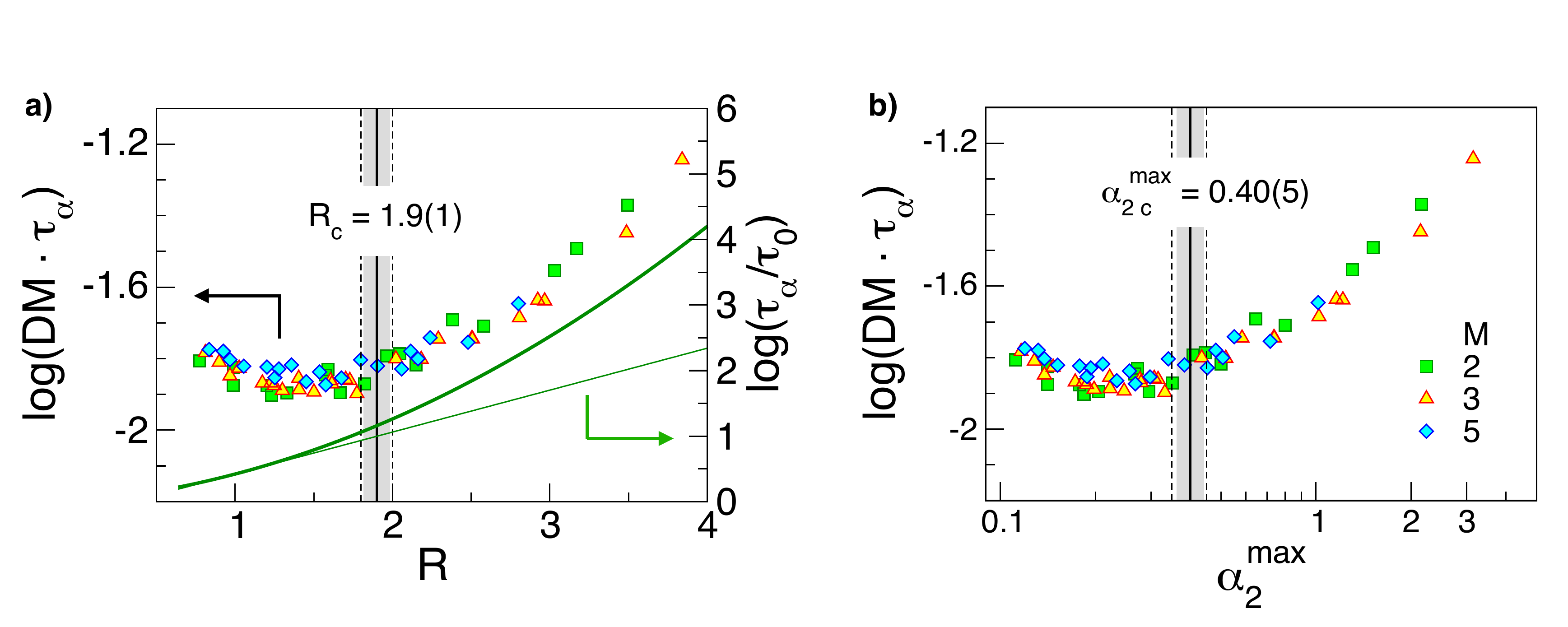}
\caption{(\textbf{a}) The~product $D\, M \, \tau_{\alpha}$ vs. the ratio $R$, Equation~\eqref{DefR2} (\textbf{b}) the same product vs. $\alpha_{2}^{max}$, the~maximum of the non-Gaussian parameter, Equation~\eqref{Eq:alpha2}. 
The~onset of the Stokes--Einstein (SE) violation  for  $\alpha_{2}^{max} > \alpha_{2 , c}^{max}$ and $R > R_{c}$, respectively, is  indicated with the full vertical lines (uncertainty marked by dashed lines). The~thick line in the panel (\textbf{a}) is the master curve between $\log \tau_\alpha$ and the DW factor, Equation~\eqref{parabolaMD}, recast in terms of $R$ and the thin line, is the corresponding linear approximation for small $R$ values. Note that the SE violation is apparent where the linear approximation is poor. Data from~\cite{Puosi12SE}.}
\label{fig3JCP12SE}
\end{figure}

Figure~\ref{fig3JCP12SE} shows that in states with homogeneous Gaussian dynamics, i.e., with small $\alpha_2^{\, max}$ values, the $R$ values are comparable or less than the unit value and the product $D\, M \, \tau_{\alpha}$ is nearly constant, i.e., the SE law holds true. {On the other hand, in the presence of significant DH, i.e., $\alpha_2^{\, max} > \alpha_{2 , c}^{max} = 0.40(5)$, one finds $R > R_c = 1.9(1)$ and the product $D\, M \, \tau_{\alpha}$ tends to increase, i.e., the SE law fails~\cite{Ediger00,Richert02,CristianoSE}}. 
The~comparison between  $\alpha_2^{\, max}$ and $R$ substantiates the conclusion that the magnitude of the ratio $R$ allows one to conclude whether DH is appreciable or not, as suggested by Equation~\eqref{Eq:GaussianDistro}. As the ratio $R$---apart from constants---depends only on DW, see Equation~\eqref{DefR2}, the finding supports previous conclusions that the long-time DH is rooted in the fast dynamics~\cite{Harrowell06}.\vspace{-6pt}

It is seen that states with equal $R$ (Figure~\ref{fig3JCP12SE}a), i.e., states with equal DW factor according to Equation~\eqref{DefR2}, exhibit nearly equal values of the product $D\, M \, \tau_{\alpha}$. A similar result has been reported for atomic binary mixtures~\cite{SpecialIssueJCP13} and metallic alloys~\cite{Puosi18SE}.  Recognising that the diffusivity $D$ and the structural relaxation time $\tau_\alpha$ reflect the long time behaviour of MSD and ISF, respectively, Figure~\ref{fig3JCP12SE}a reveals that Equation~\eqref{u2X} is valid even in the {\it diffusive regime} which is entered in polymer melts at times fairly longer than $\tau_{ee}$, being $\tau_{ee} \gg \tau_{\alpha}$.

\subsection{Quasi-Universal SE Breakdown of Fragile Glass-Formers}
\label{QuasiUnivSE}

Having noted that the SE failure is tracked by the DW factor in unentangled polymers, we now pose the question if this finding exhibits universal features. To this aim, we consider  the ratio $K_{SE}/K_0$ with $K_{SE}$ defined in Equation~\eqref{SEProduct} and $K_0$  a scaling factor to ensure the unit limit value at large DW~factor. 

\begin{figure}[H]
\centering
\includegraphics[width=8 cm]{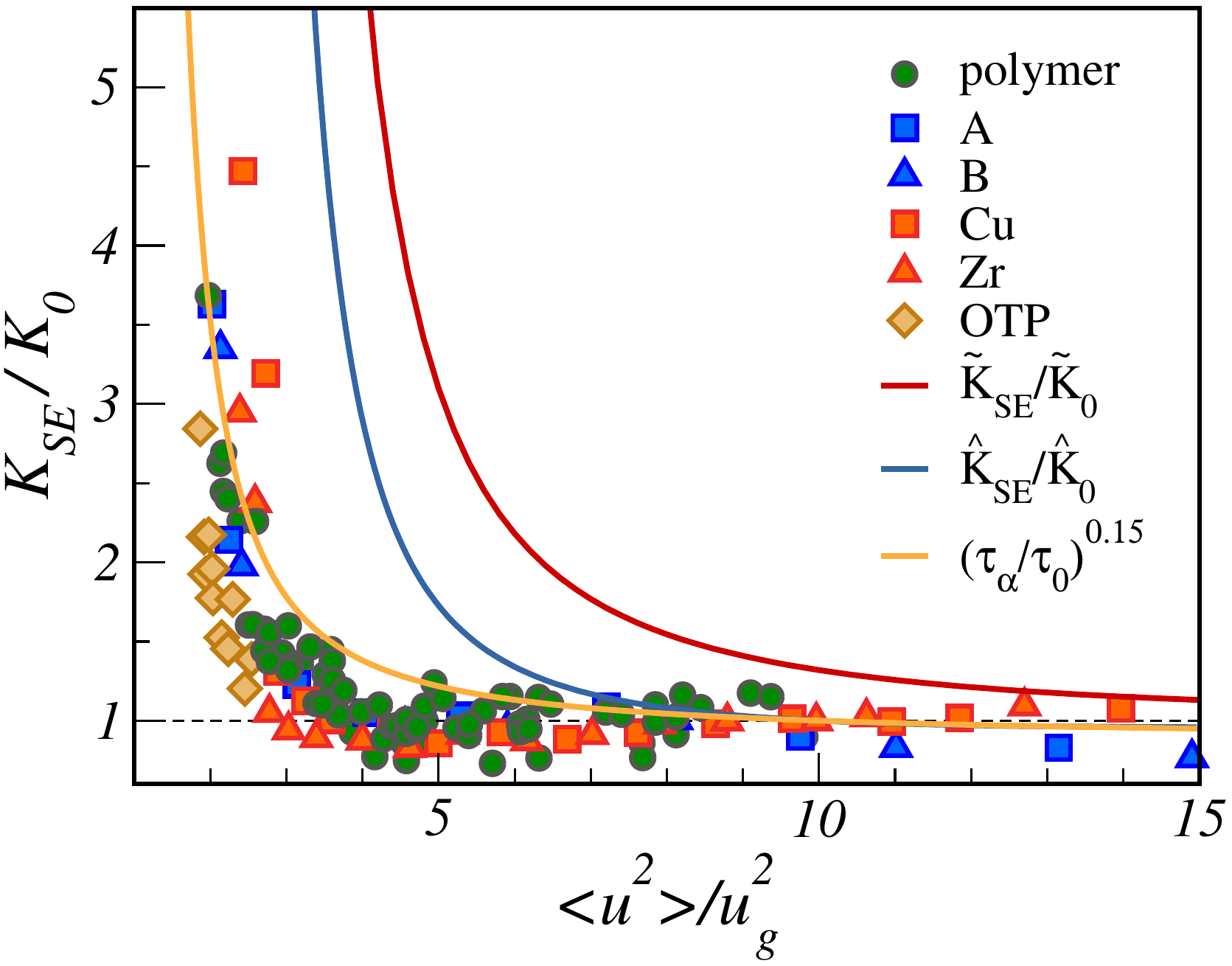}
\caption{Stokes--Einstein product $K_{SE}$, normalised by its high temperature value $K_0$ ($\tau_\alpha \simeq 1$ ps), as a function of the reduced DW factor  $\langle u^2 \rangle/u^2_g$, $ u^2_g$ being the DW factor at the glass transition. In~addition to unentangled polymers, the plot also considers MD data concerning atomic binary mixtures (atoms labelled as $A$ and $B$) and metallic alloys made by Cu and Zr atoms, as well as experimental data for ortho-terphenyl (OTP)~\cite{tolleRepProg01,SillescuJPCB97}. Two predictions of the master curve are presented in terms of the quantity $\hat{K}_{SE}$ and $\tilde{K}_{SE}$, Equations~\eqref{K_Hat} and  \eqref{K_TildeVsZ}, respectively. Both quantities have no adjustable parameters. $\hat{K}_{0}$ and $\tilde{K}_{0}$ are suitable constants to ensure the unit limit value at large $\langle u^2 \rangle/u^2_g$. A third master curve, drawn from the fractional SE law $\tau_\alpha ^{1-\kappa}$ with $\kappa = 0.85$ (orange  curve), is superimposed to the other curves. For numerical data, $u^2_g$ is obtained according to the procedure outlined in~\cite{OurNatPhys}. Data from \cite{Puosi18SE}.}
\label{fig4JCP18}
\end{figure}

In Figure~\ref{fig4JCP18}, we plot the ratio $K_{SE}/K_0$ 
 as a function of  $\langle u^2 \rangle/u^2_g$. We complement the MD results on unentangled polymers already presented in Figure~\ref{fig3JCP12SE} with other MD data, considering the diffusion of  Cu and Zr atoms in metallic alloy, A and B atoms in a Lennard--Jones binary mixture~\cite{Puosi18SE} together with experimental data concerning the popular fragile glass-former ortho-terphenyl (OTP)~\cite{tolleRepProg01,SillescuJPCB97}.  Figure~\ref{fig4JCP18} evidences the good collapse of the SE violation in terms of the reduced DW.
Figure~\ref{fig4JCP18} offers the opportunity to test the master curve predicted by the model of Section~\ref{ModelDW}  with  no adjustable parameters $\hat{K}_{SE}$, Equation~\eqref{K_Hat}, and its approximation $\tilde{K}_{SE}$, Equation~\eqref{K_TildeVsZ}. 
It is seen that $\hat{K}_{SE}$ predicts a stronger SE breakdown than actually observed. Larger deviations are exhibited by the approximant $\tilde{K}_{SE}$.
How to improve the agreement: The~expression of the diffusion coefficient in Section~\ref{diffcoeff}, assuming that displacements as large as $a$ are statistically independent, aims at a SE product $\hat{K}_{SE}$, Equation~\eqref{K_Hat}, with no additional adjustable parameters with respect to the ones of $\tau_\alpha$, i.e., the ones of Equation~\eqref{lnscaledparabola}. This puts severe constraints on the shape of the distribution of the square displacements needed to overcome the relevant energy barriers $p(a^2)$, Equation~\eqref{Eq:GaussianDistro2}. The~form of the distribution is adequate for  large displacements {to reach the transition state} governing $\tau_\alpha$, as proven by the effective fit of the MD data by the predicted master curve shown in Figure~\ref{fig7JCP09}. However,
the~findings of Figure~\ref{fig4JCP18} suggest that it must be improved for  small displacements affecting $D$. Alternatively,  we may also state that the distribution $p(\ln\tau)$, Equation~\eqref{lognormal}, should be refined as far as the short relaxation times are concerned.

\textls[-19]{Figure~\ref{fig4JCP18} shows that better agreement occurs by assuming the fractional SE form  $D \tau_\alpha \simeq  \tau_\alpha ^{1-\kappa}$~\mbox{\cite{SillescuSEJNCS94,DouglasLepoJNCS98}}} with $\tau_\alpha$ as given from Equation~\eqref{lnscaledparabola}. The~best fit is found with $\kappa = 0.85$, which equals the universal value found by Mallamace et al.~\cite{MallamaceFSEPNAS10}, deviating from the prediction of the ``obstruction model'' \mbox{$\kappa=2/3$}~\cite{DouglasLepoJNCS98}.

\section{Displacement Correlation Length}
\label{DisplDisplCorFun}

Several results of the present paper suggest that the vibrational dynamics, as sensed by DW factor, provides insight into DH. A line of attack to understand how vibrational dynamics is related to slow relaxation is provided by the model of Section~\ref{ModelDW}. The~model is based on the
 distribution of the (squared) displacements needed by a particle to rearrange in the different local environments, Equation~\eqref{Eq:GaussianDistro2}, leading in turn to the distribution of relaxation times, Equation~\eqref{lognormal}. As noted in Section~\ref{QuasiUnivSE},  the model needs further development. A further aspect to be improved is the absence of any detail on the localization of the particles with a given dynamics. This prevents any prediction concerning
a peculiar aspect of DH, i.e., the existence of spatial domains with characteristic length scales where the particles undergo correlated motion, e.g., see Figure~\ref{HOM_HET}. 

To make progress, it is worthwhile to preliminarily judge whether DW exhibits some correlation with possible dynamic length scales. 
To pursue this task, we studied the following monomer displacement--displacement correlation (DDC) functions~\cite{PuosiLepoJCPCor12,PuosiLepoJCPCor12_Erratum}:
\begin{eqnarray}
C_{\vec{u}}(r,\tau_\alpha)&=&\langle \hat{\bf u}_i(t_0,\tau_\alpha) \cdot \hat{\bf u}_j (t_0,\tau_\alpha) \rangle, \\
C_{\delta u}(r,\tau_\alpha) &=&\langle \delta u_i (t_0,\tau_\alpha) \delta u_j (t_0,\tau_\alpha) \rangle/\langle [\delta u(t_0,\tau_\alpha)]^2 \rangle.
\end{eqnarray}

An average over all the $i$-th and $j$-th monomers spaced by $r$ is understood.  $ \hat{\bf u}_k (t_0,t)$ is the versor of the displacement vector of k-th monomer in a time interval from $t_0$ to $t_0+t$,  $ {\bf u}_k (t_0,t)={\bf r}_k(t_0+t) - {\bf r}_k (t_0)$ and $\delta u_k(t_0,t)=| {\bf u}_k (t_0,t)|- \langle | {\bf u}(t_0,t)| \rangle$, where $| {\bf u}_k (t_0,t)|$ is the modulus of the displacement. Henceforth, $C_{\delta u}(r,\tau_\alpha)$ and $C_{\vec{u}}(r,\tau_\alpha)$ will be referred to as modulus (or mobility) and direction DDC functions, respectively.
{ 
Local anisotropies and collective elastic solid-like response to the rattling of the monomers in the cage of their neighbours play a central role in the DDC build-up~\cite{BerniniCageEffectJCP16}.}

We  consider DDCs of the states presented in part of the states in Figure~\ref{fig1JCP09}. We remind that the states (i) exhibit different DH degree, e.g., see Figures~\ref{fig4JPCB11fig5JCP12} and  \ref{fig2JCP12SE}, and (ii) are grouped 
in sets labelled B through E, each set being characterized by a single value of the DW factor.

\begin{figure}[H]
\centering
\includegraphics[width=14 cm]{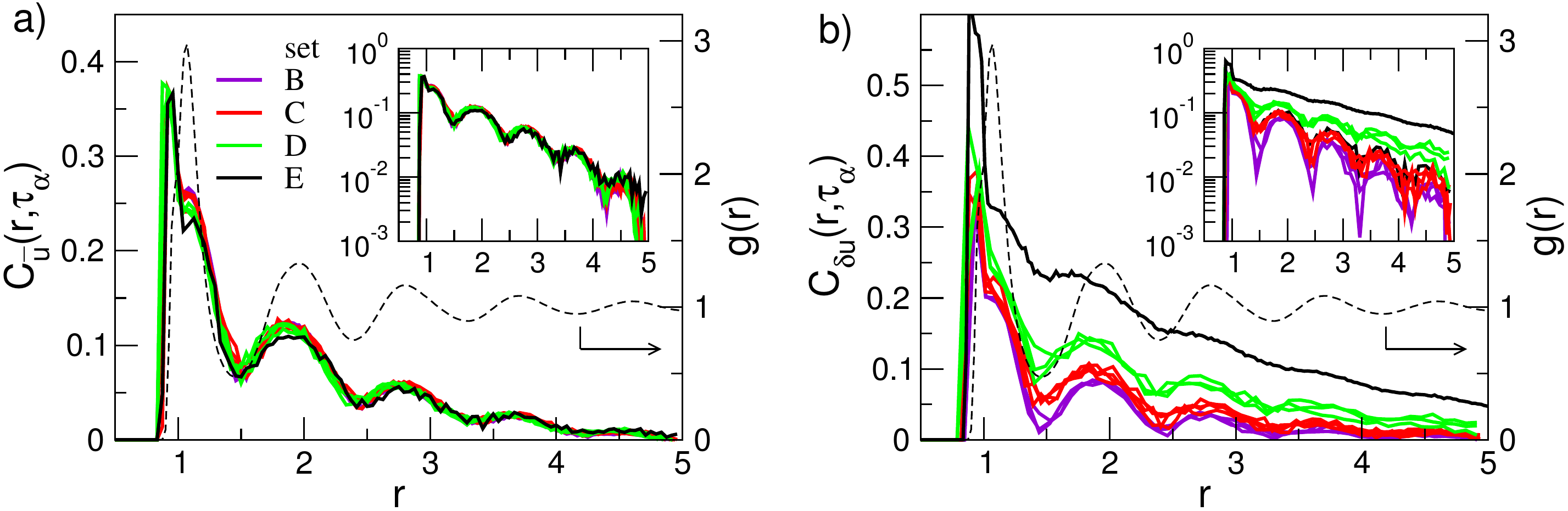}
\caption{Radial dependence of the correlation of the direction (\textbf{a}) and the mobility   (\textbf{b})  displacements occurring in a time range as wide as the structural relaxation time $\tau_\alpha$. For comparison, the radial distribution function $g(r)$ (dashed line) of the state  with $\{M=2, \rho =1.086,T=0.7,q=7,p=6\}$ is plotted. { Note that $g(r)$ is virtually state-independent.} The~insets are semi-log plots of the corresponding main panels. Note the approximate exponential decay of the peak amplitudes with slopes $\xi_{\vec{u}}(\tau_\alpha)$ and $\xi_{\delta u}(\tau_\alpha)$, respectively.  Data from \cite{PuosiLepoJCPCor12}.}
\label{figcor}
\end{figure}

\textls[-15]{Figure~\ref{figcor}a,b shows the spatial dependence of the direction and the modulus DDC functions, respectively, for  the sets of states labelled B through E in Figure~\ref{fig1JCP09}.} Both correlation functions manifest damped oscillations in-phase with the pair correlation function $g(r)$, thus evidencing that the correlated motion of a tagged monomer and its surroundings is influenced by the structure of the latter. This agrees with previous work on DDCs in Lennard--Jones systems~\cite{BennemannNat99,DonatiGlotzerPRL99}, hard-sphere~\cite{DoliwaPRE00} and experiments on colloids~\cite{WeeksJPCM07}. The~highest correlations are reached at a distance corresponding to the bond length $b=0.97$ which demonstrates the highly concerted dynamics of bonded monomers. The~correlation peaks, located at the first-, second-,... neighbours shells, vanish approximately in an exponential way on increasing the distance from the tagged particle (see insets of Figure~\ref{figcor}). In more detail, Figure~\ref{figcor}a shows that the direction correlations do not show significant increase in their spatial extension on increasing the structural relaxation time. Figure~\ref{figcor}b shows the modulus (mobility) correlations. Differently from the direction correlations, their spatial extension  increase meaningfully with the structural relaxation time (see also the inset of Figure~\ref{figcor}b).

\begin{figure}[H]
\centering
\includegraphics[width=8 cm]{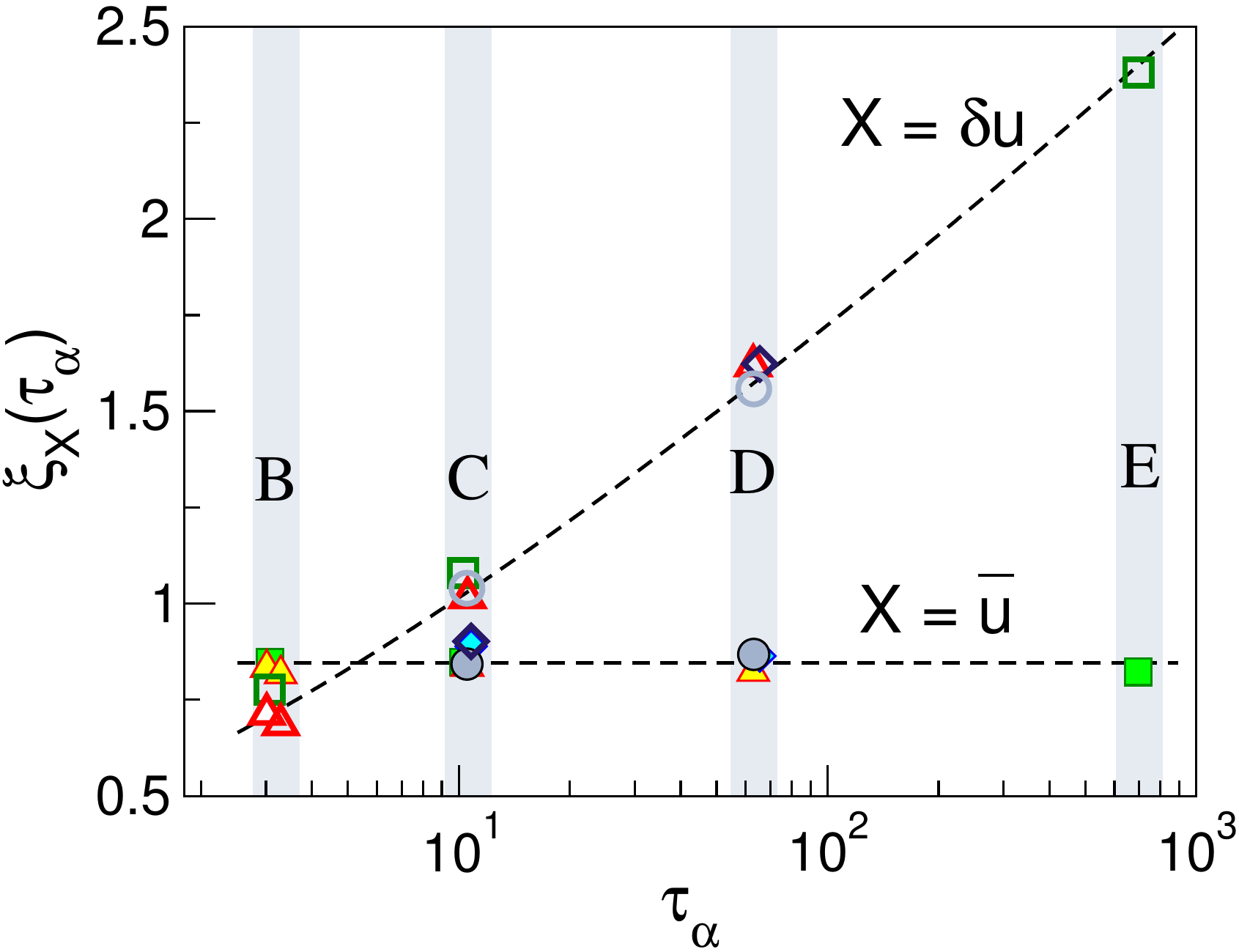}
\caption{The~direction $\xi_{\vec{u}}(\tau_\alpha)$ (full symbols) and modulus $\xi_{\delta u}(\tau_\alpha)$ (open symbols) correlation lengths vs. the structural relaxation time $\tau_\alpha$ of selected set of states of Figure~\ref{fig1JCP09}. Dashed lines are guides for the eyes. States with  equal DW factor, i.e., belonging to the same set B, $\cdots$, E exhibit  equal directional and mobility correlation lengths. Data from \cite{PuosiLepoJCPCor12}.}
\label{CorrLength}
\end{figure}

Figure~\ref{figcor}b clearly shows that physical states with equal DW factor, i.e., belonging to the same set of states (B, . . . , E), exhibit the same spatial correlations. { This provides further support that Equation~\eqref{u2X} also holds if the spatial dependence of the correlation function is considered for a given time up to $\tau_\alpha$ at least.} To provide additional insight, we evaluated the length scales of the exponential decays of the DDC maxima with the distance $\sim \exp [-r/\xi_X(\tau_\alpha)]$ with $X = \vec{u}, \delta u$, thus defining two distinct dynamic correlation lengths pertaining to direction and modulus DDCs, $\xi_{\vec{u}}(\tau_\alpha)$ and $\xi_{\delta u}(\tau_\alpha)$, respectively. Figure~\ref{CorrLength} shows these quantities. It is seen that the spatial extension of the modulus DDC 
 increases quite a lot  with $\tau_\alpha$ and reaches distances beyond the next-nearest neighbours for the states with the slowest relaxation. Instead, the direction correlations are virtually independent of the structural relaxation. Irrespective of the correlation length under consideration, Figure~\ref{CorrLength}  also shows that they are  equal within the errors for states with equal DW factor, i.e., belonging to the same set of states (B, . . . , E).
We are now in a position to compare our results with previous work on DDCs. Simulations of Lennard--Jones binary mixture (BM) observed that at  time $t_\alpha$, corresponding to  maximum dynamic heterogeneity, $\xi^{BM}_{\delta u}(t_\alpha)$ increases as the temperature decreases, whereas $\xi^{BM}_{\vec{u}}(t_\alpha)$ is almost constant~\cite{TokuyamaPhilMag}. This agrees with our findings in Figure~\ref{CorrLength} concerning unentangled polymers.
As to the modulus DDC correlation length, one finds~\cite{PuosiLepoJCPCor12} that after suitable algebraic manipulation  to allow comparison~\cite{WeeksJPCM07}, our changes of $\xi_{\delta u}(\tau_\alpha)$ with $\tau_\alpha$ are in quantitative agreement with the results of Bennemann et al. reported in a study of the same polymer system investigated here~\cite{BennemannNat99}.

\section{Discussion}

There is wide experimental, numerical and theoretical evidence that the fast vibrational dynamics, as sensed by the Debye--Waller factor $\langle u^2\rangle$,  and the time scale $\tau_\alpha$ of the slow microscopic reorganisation of a liquid close to the transition to the glassy state are correlated in an universal way. Potential applicative implications concerning the quick characterisation of the stability of disordered structures with ultraslow relaxation are apparent.  Less attention has been paid to a series of numerical MD simulation studies concluding in favour of strong correlations also between the vibrational dynamics and the dynamical heterogeneity, the spatial distribution of long-time mobility developing when approaching the disordered solid state. 
{ We reviewed these studies, mainly concerning melts of unentangled linear polymers, unifying all the results for the first time in terms of Equation~\eqref{u2X}. The~latter has been tested both in  space and time. In particular, we considered time-dependent quantities accounting for transport and relaxation like MSD, ISF and NGP and showed that they are related to the self-part of the van Hove function, which reduces to suitable correlation functions. In this respect, the correlation between the Debye--Waller factor and the breakdown of the SE law, a hallmark of DH presence, is seen as an ancillary consequence of the extension of Equation~\eqref{u2X}, which at times is much longer than $\tau_\alpha$, where the motion is diffusive.  We also inspected Equation~\eqref{u2X} in space by considering both the van Hove function and DDC functions. Notably, DDC functions are  collective in nature, differently from 
the self-part of the van Hove function, which is a  single-particle quantity. This suggests that Equation~\eqref{u2X} also holds for  collective correlation functions. A further validation of this conclusion is offered by the collective stress--stress correlation function, which has been presented elsewhere~\cite{Puosi12} and not discussed in this review.}

The~understanding of the microscopic origin of the correlation between the vibrational dynamics and the heterogeneous dynamics close to the glass transition is still unsatisfactory in several respects. In particular, both the model discussed here, as well as other ones reported in the literature, even if successful in relating the Debye--Waller factor $\langle u^2\rangle$ with the time scale $\tau_\alpha$, are currently unable to account for the fact evidenced by the numerical simulations that the vibrational dynamics conveys also information on the spatial correlations between the mobility of different particles.

\section{Methods}
\label{method} 

Most results discussed in this review concern a coarse-grained model of a linear polymer chain with $M$ monomers is adopted. Bending and torsional potentials are neglected, i.e., the chain is fully flexible. While addressing the interested reader to the referenced papers for further details, we provide here some general aspect of the numerical model. We considered systems with total number of monomer~$N \ge2000$. 
Non-bonded monomers at a distance $r$ interact via the truncated parametric potential:
\begin{equation}
U_{q,p}(r) = \frac{\epsilon}{p-q} \left [  p \left (\frac{\sigma^*}{r}\right )^{q} - q \left (\frac{\sigma^*}{r}\right )^{p}\right ] + U_{cut}
\label{Eq:modifiedLJ}
\end{equation}
where $\sigma^* = 2^{1/6} \sigma$ and the value of the constant $U_{cut}$ are chosen to ensure $U_{p,q}(r) = 0$ at $r \ge r_c = 2.5 \sigma $. The~minimum of the potential $U_{p,q}(r)$ is at 
$r = \sigma^*$, with a constant depth $U(r = \sigma^*) = \epsilon$. Note that $U_{q,p}(r) = U_{p,q}(r)$. 
Bonded monomers interact with a potential which is the sum of the Finitely Extendible Nonlinear Elastic (FENE) potential and the Lennard--Jones (LJ) potential~\cite{sim}. The~resulting bond length is $b = 0.97\sigma$, within a few percent.
We set $\sigma = 1$ and $\epsilon = 1$. The~time unit is $\tau_{MD} = (m\sigma^2 / \epsilon)^{1/2}$, with $m$ being the mass of the monomer. 
Temperature is in units of $\epsilon/k_B$, where $k_B$ is the
Boltzmann constant. We set $m = k_B = 1$. {All the data presented in this work are expressed in reduced MD units}.
It is interesting to map the reduced MD units to  real physical units. The~procedure involves the comparison of the experiment with simulations and provide the basic length ($\sigma$), temperature ($\epsilon/ k_B$) and time ($\tau_{MD}$) units~\cite{KremerGrestJCP90,Kroger04, PaulSmith04,SommerLuoCompPhysComm09,sim}. For polyethylene  and polystyrene,  it was found \mbox{$\sigma = 5.3$~\AA}, $\epsilon/ k_B = 443$~K, \mbox{$\tau_{MD} = 1.8$~ps}  and $\sigma = 9.7 $ \AA, $\epsilon/ k_B = 490$~K, \mbox{$\tau_{MD} = 9$ ps,} respectively~\cite{Kroger04}. For~poly(vinyl
alcohol) \mbox{$\sigma = 5.2$ \AA}, $\epsilon/ k_B = 550$~K and $\tau_{MD} = 1.63$ ps~\cite{SommerLuoCompPhysComm09}. For polyisoprene $\sigma = 6.7$ \AA, $\epsilon/ k_B = 307$~K and \mbox{$\tau_{MD} = 10$ ps ~\cite{KremerGrestJCP90}}. The~densities used in this and other studies are lower than the densities at atmospheric pressure, e.g., when mapping our model to  polyethylene  and polystyrene, one finds $ \sim $0.5 and $ \sim $0.7 g/cm${}^3$, to be compared to the actual values $0.78$ and $0.92$~g/cm${}^3$, respectively~\cite{Kroger04}.

\vspace{6pt} 



\authorcontributions{F.P., A.T. and D.L. wrote the manuscript together. All authors have read and approved the final version.}

\funding{This research was funded by the project PRA- 2018-34 (``ANISE'') from the University of Pisa. }

\acknowledgments{A generous grant of computing time from IT Center, University of Pisa, and Dell EMC\textsuperscript{\textregistered}  Italia is gratefully acknowledged.}

\conflictsofinterest{The~authors declare no conflicts of interest.} 

\abbreviations{The~following abbreviations are used in this manuscript.\\

\noindent 
\begin{tabular}{@{}ll}
DDC & displacement--displacement correlation \\
DH & dynamical heterogeneity \\
DW & Debye--Waller \\
ISF & Intermediate scattering function\\
MD & Molecular-dynamics\\
MSD & Mean square displacement\\
NGP & non-Gaussian parameter \\
SE & Stokes--Einstein
\end{tabular}}

\appendixtitles{no} 
\appendix
\section{}\label{app1}
\unskip\appendixtitles{yes}
\subsection{Structural Relaxation}
\label{appendixStrucRelax}

According to the model detailed in Section~\ref{relaxtime}, the master curve relating the structural relaxation time $\tau_\alpha$ and the DW factor $\langle u^2\rangle$, Equation~\eqref{eqn:u2tauExpGeneric}, takes the form given by Equation~\eqref{parabola2}. Other variants of Equation~\eqref{parabola2} are of interest in the comparison with numerical and experimental results. As an example, the best fit of the master curve Equation~\eqref{parabola2} with the numerical data concerning the melt of unentangled polymer chains of interest here yields (in MD units)~\cite{OurNatPhys,lepoJCP09}:
\begin{equation}
\log \tau_\alpha = \alpha + \beta \frac{1}{\langle u^2\rangle} + \gamma \frac{1}{\langle u^2\rangle^2}
\label{parabolaMD}
\end{equation}
where $\alpha =-0.424(1), \beta = \overline{a^2}/(2 \ln10) = 2.7(1) \cdot 10^{-2}, \gamma = \sigma^2_{a^2}/(8 \ln10) =  3.41(3) \cdot 10^{-3}$. 

To recast Equation~\eqref{parabolaMD} as a  universal master curve removing system-dependent quantities, one considers
the DW factor at the glass transition $u^2_g$ (defined via $\tau_\alpha = 10^2\, \mbox{s}$) and introduces the reduced variable $\langle u^2 \rangle/ \langle u^2_g \rangle$, so as to write Equation~\eqref{parabolaMD}, 
\begin{equation}
\ln \tau_\alpha = \hat{\alpha}  + \hat{\beta} \frac{\langle u^2_g \rangle}{\langle u^2 \rangle} +\hat{\gamma} \left (\frac{\langle u^2_g \rangle}{\langle u^2\rangle} \right )^2
\label{lnscaledparabola}
\end{equation}
where $ \hat{\alpha} = 2 \ln 10  - \hat{\beta} - \hat{\gamma}$.  The~ansatz is that Equation~\eqref{lnscaledparabola} is system-independent and both $\hat{\beta}$ and $\hat{\gamma}$ are universal coefficients. To derive their numerical values, we use the value $< u^2_g >^{1/2} = 0.129$ and the best-fit values of $\alpha,\beta,\gamma$, drawn from the numerical simulations of the melt of unentangled polymer chains~\cite{OurNatPhys,lepoJCP09}. This yields $\hat{\beta}=3.7(1)$ and $\hat{\gamma}=28.4(2)$~\cite{Puosi18SE}.

\subsection{Diffusion Coefficient}
\label{appendixDiff}

The~diffusion coefficient is evaluated from Equation~\eqref{diffusion} via Equations~\eqref{dyreWolynes} and  \eqref{Eq:GaussianDistro2}. The~result is
\vspace{12pt}
\begin{equation}
D (\langle u^2\rangle) =  \frac{\sigma_{a^2}}{6 \tau_0} \exp\left [ - \frac{1}{2}\left (\frac{\overline{a^2}}{\sigma_{a^2}} \right )^2 \right ] 
\frac{G(\langle u^2\rangle)}{ \sqrt{18 \pi} \left [1+ \textrm{erf}(\overline{a^2}/ \sqrt{2} \sigma_{a^2}) \right ]}
\label{diffusionEXPL}
\end{equation}
where
\begin{equation}
\left \{ 
\begin{array}{l}
G(x) = \left \{ 1 - \sqrt{\pi} \, \Lambda_{-}(x) \exp \left [\Lambda^2_{-}(x) \right] \textrm{erfc}[\Lambda_{-}(x)] \right \}  \vspace{2mm} \\
\Lambda_{\pm}(x) = \frac{\sigma_{a^2}^2 \pm 2 \, \overline{a^2} \, x }{\sqrt{8} \, \sigma_{a^2} x} \\
\end{array}.
\right.
\end{equation}

As  $ \tau_\alpha^{(HW)}(a^2)$ depends on the parameter $a$ in a much more marked way than $a^2$, see Equation~\eqref{dyreWolynes},  an effective approximation of the diffusivity, as expressed by Equation~\eqref{diffusion},  is $D \simeq \tilde{D}$ with
\begin{equation}
\tilde{D} = \frac{a^2_0}{6}  \left \langle  1/ \tau_\alpha^{(HW)}(a^2) \right \rangle_{a^2}
\end{equation}
where $a_0$ is a constant.

\subsection{Stokes--Einstein Product}
\label{appendixSE}

After suitable manipulation the Stokes--Einstein product $K_{SE}$, Equation~\eqref{SEProduct}, takes the from ($M=1$)
\begin{equation}
K_{SE}(\langle u^2\rangle) = D (\langle u^2\rangle) \, \times \, \tau_\alpha(\langle u^2\rangle)
\label{KSE}
\end{equation}
where
\begin{equation}
\left \{ 
\begin{array}{l}
K_{SE}(x) = \sigma_{a^2} \, F(x) G(x) \vspace{2mm} \\
F(x) = \frac{ 1 }{\sqrt{18 \pi}}
\exp \left [\Lambda^2_{+}(x) \right ]
\exp\left [ - \left (\frac{\overline{a^2}}{\sigma_{a^2}} \right )^2 \right ]
\frac{1+\textrm{erf}(\Lambda_{+}[x])}{\left [1+ \textrm{erf}(\overline{a^2}/ \sqrt{2} \sigma_{a^2}) \right ]^2}
\end{array}.
\right.
\end{equation}
where the auxiliary functions $G(x)$ and $\Lambda_{\pm}(x)$ are defined in  Appendix~\ref{appendixDiff}.

When expressed in terms of the adimensional quantity $z =\langle u^2 \rangle/u^2_g$, the product $K_{SE}$ takes the form $\hat{K}_{SE}(z)$ with
\begin{equation}
\left \{ 
\begin{array}{l}
\hat{K}_{SE}(z) =  u^2_g \, \hat{F}(z) \left \{ 1 - \sqrt{\pi} \, \hat{\Lambda}_{-}(z) \exp \left [\hat{\Lambda}^2_{-}(z) \right] \textrm{erfc}[\hat{\Lambda}_{-}(z)] \right \} \vspace{2mm} \\
\hat{\Lambda}_{\pm}(z) = \frac{2 \hat{\gamma} \pm \hat{\beta} z }{2 \hat{\gamma}^{1/2} z } \vspace{2mm}\\
\hat{F}(z) =  \frac{2}{3} \sqrt{\frac{\hat{\gamma}}{\pi}}
\exp \left [\hat{\Lambda}^2_{+}(z) \right ]
\exp\left [ - \frac{\hat{\beta}^2}{2 \hat{\gamma}} \right ]
\frac{1+\textrm{erf}(\hat{\Lambda}_{+}[z])}{\left [1+ \textrm{erf}(\hat{\beta}/ 2 {\hat{\gamma}}^{1/2})\right ]^2}
\end{array}.
\right.
\label{K_Hat}
\end{equation}

The~quantities $\hat{\beta}$ and $\hat{\gamma}$ are defined in Appendix~\ref{appendixStrucRelax}. 

The~Stokes--Einstein product $K_{SE}$ may be approximated as $K_{SE} \simeq \tilde{K}_{SE}$ with $\tilde{K}_{SE} = \tilde{D}\tau_\alpha$, where  $\tilde{D}$ is defined in in  Appendix~\ref{appendixDiff}. This yields
\begin{equation}
\tilde{K}_{SE} = \frac{a^2_0}{6}  \left \langle  \tau_\alpha^{(HW)}(a^2) \right \rangle_{a^2} \left \langle  1/ \tau_\alpha^{(HW)}(a^2) \right \rangle_{a^2}
\label{K_Tilde}
\end{equation}

The~explicit expression of the quantity $\tilde{K}_{SE}$, in terms of $z= \langle u^2 \rangle/u^2_g$, reads
\begin{equation}\label{K_TildeVsZ}
\tilde{K}_{SE}(z) = \frac{a^2_0}{6} \exp \left [  2 \hat{\gamma}/z^{2} \right ]   \frac{ \left [ 1+\mbox{erf} \left ( \frac{\hat{\gamma}^{1/2}}{z} + \frac{\hat{\beta}}{2\hat{\gamma}^{1/2}}  \right ) \right] \left [ \mbox{erfc}  \left ( \frac{\hat{\gamma}^{1/2}}{z} - \frac{\hat{\beta}}{2\hat{\gamma}^{1/2}}  \right ) \right]}{\left [  1+ \mbox{erf} \left ( \frac{\hat{\beta}}{2\hat{\gamma}^{1/2}} \right) \right]^2} 
\end{equation}
\appendixtitles{no}
\section{}
\label{appendixvanHove}

The~van Hove function for a uniform fluid is defined as
\begin{equation}
G({\bf r},t) =  \frac{1}{N} \langle   \sum_{i=1}^N \sum_{j=1}^N  \delta [{\bf r}+{\bf x}_i(0)-{\bf x}_j(t)] \rangle
\label{vanhove0}
\end{equation}

Physically, $G d{\bf r}$ is the probability of finding a particle $j$ in a region $d{\bf r}$ around a point ${\bf r}$ at time $t$ if the particle $i$ was at the origin at time $0$. We may recast the van Hove function by resorting to the time-dependent, microscopic particle density~\cite{HansenMcDonaldIIIEd}
\begin{equation}
\rho({\bf r},t) = \sum_{i=1}^N \delta [{\bf r}-{\bf x}_i(t)]
\label{density}
\end{equation}

We rewrite Equation~\eqref{vanhove0} as 
\begin{eqnarray}
G({\bf r},t) &=& \frac{1}{N}  \langle \int \sum_{i=1}^N \sum_{j=1}^N \delta [{\bf r}'+ {\bf r}-{\bf x}_j(t)] \delta[{\bf r}'-{\bf x}_i(0)] d{\bf r}' \rangle \label{passage1}  \\
&=& \frac{1}{N} \int  \langle \rho({\bf r}'+ {\bf r},t) \rho({\bf r}',0) d{\bf r}' \rangle \\
&=& \frac{1}{\rho}  \langle \rho({\bf r},t) \rho({\bf 0},0) \rangle \label{corVH}
\end{eqnarray}
where $\rho$ is the average number density. Equation~\eqref{corVH} shows that the van Hove function is proportional to the density correlation function. It is easily shown that the van Hove function may be written as~\cite{HansenMcDonaldIIIEd}
\begin{equation}
G({\bf r},t) = G_{s}({\bf r},t) + G_{d}({\bf r},t) 
\end{equation}
where $G_{s}({\bf r},t)$ and $G_{d}({\bf r},t)$ are usually called the ``self'' and ``distinct'' parts. Equation~\eqref{vanhove} provides the explicit expression of $G_{s}({\bf r},t)$. The~distinct part is written as
\begin{equation}
G_d({\bf r},t) =  \frac{1}{N} \langle   \sum_{i=1}^N \sum_{j\neq i}^N  \delta [{\bf r}+{\bf x}_i(0)-{\bf x}_j(t)] \rangle
\label{vanhoved}
\end{equation}

Finally, we show that both $G_{s}({\bf r},t)$ and $G_{d}({\bf r},t)$ may be expressed in terms of suitable correlation functions. To this aim, we define the auxiliary function $B^i({\bf r},t) \equiv \delta [{\bf r}-{\bf x}_i(t)]$. By repeating the same passages leading from Equation~\eqref{vanhove0} to Equation~\eqref{passage1}, one has 
\begin{eqnarray}
G_s({\bf r},t) &=& \frac{1}{N}  \langle \int \sum_{i=1}^N \delta [{\bf r}'+ {\bf r}-{\bf x}_i(t)] \delta[{\bf r}'-{\bf x}_i(0)] d{\bf r}' \rangle  \\
&=& \frac{1}{N} \int \sum_{i=1}^N  \langle B^i({\bf r}'+{\bf r},t) B^i({\bf r}',0) d{\bf r}' \rangle \\
&=& \frac{1}{\rho} \sum_{i=1}^N  \langle B^i({\bf r},t) B^i({\bf 0},0) \rangle 
\end{eqnarray}

The~last passage follows from the uniformity of the fluid. Analogously, one finds
\begin{equation}
G_d({\bf r},t) =  \frac{1}{\rho} \sum_{i=1}^N \sum_{j\neq i}^N  \langle B^i({\bf r},t) B^j({\bf 0},0) \rangle 
\end{equation}
%

\reftitle{References}

\end{document}